\documentclass[a4paper,11pt]{article}
\pdfoutput=1 
\usepackage{jcappub}
\usepackage[T1]{fontenc}
\author[a]{Asta~Heinesen} 

\affiliation[a]{Univ Lyon, Ens de Lyon, Univ Lyon1, CNRS, Centre de Recherche Astrophysique de Lyon UMR5574, F--69007, Lyon, France}

\emailAdd{asta.heinesen@ens--lyon.fr}

\usepackage{changes}
\usepackage{subcaption}
\usepackage[section]{placeins}
\usepackage[titletoc]{appendix}
\usepackage{xcolor}
\usepackage{amssymb}
\usepackage{dsfont}
\usepackage{bm}
\usepackage{mathtools}
\usepackage{enumerate}
\DeclareMathAlphabet{\mathpzc}{OT1}{pzc}{m}{it}

\newcommand{\sayy}[1]{`#1'}
\DeclarePairedDelimiter\abs{\lvert}{\rvert}%
\providecommand{\href}[2]{#2}

\def\be{\begin{equation}}
\def\ee{\end{equation}}
\def\bea{\begin{eqnarray}}
\def\eea{\end{eqnarray}}

\def\hth{\hat{\theta}}
\def\sig{\sigma}
\def\hsig{\hat{\sigma}}
\def\Om{\Omega}
\def\la{\langle}
\def\ra{\rangle}
\def\Eu{ \mathfrak{H} }
\definecolor{MyB}{rgb}{0.1,0.1,1.0}

\title{Multipole decomposition of the general luminosity distance \sayy{Hubble law} -- a new framework for observational cosmology}

\abstract{
We present the luminosity distance series expansion to third order in redshift for a general space-time with no assumption on the metric tensor or the field equations prescribing it. 
It turns out that the coefficients of this general \sayy{Hubble law} can be expressed in terms of a \emph{finite} number of physically interpretable multipole coefficients. The multipole terms can be combined into effective direction dependent parameters replacing the Hubble constant, deceleration parameter, curvature parameter, and \sayy{jerk} parameter of the Friedmann-Lema\^{\i}tre-Robertson-Walker (FLRW) class of metrics. 
Due to the finite number of multipole coefficients, the \emph{exact} anisotropic Hubble law is given by 9, 25, 61 degrees of freedom in the $\mathcal{O}(z)$, $\mathcal{O}(z^2)$, $\mathcal{O}(z^3)$ vicinity of the observer respectively, where $z\!:=\,$redshift. 
This makes possible model independent determination of dynamical degrees of freedom of the cosmic neighbourhood of the observer and direct testing of the FLRW ansatz.  
We argue that the derived multipole representation of the general Hubble law provides a new framework with broad applications in observational cosmology. 
}
\keywords{Hubble-Lema\^{\i}tre law, distance ladder, relativistic cosmology, geometrical optics, observational cosmology} 
\catcode`\@=11
\catcode`\@=12
\begin{document}
\maketitle


\tableofcontents
\newpage 

\section{Introduction}
The Hubble-Lema\^{\i}tre law \cite{Lemaitre:1927,Slipher:1917,Hubble:1929} describing the observed approximate proportional relationship $z\! = \! H_0 d_L$ between cosmological redshift $z$ and luminosity distance $d_L$ in terms of the Hubble parameter $H_0$ is one of the most groundbreaking discoveries made in cosmology. 
The empirical discovery of positive correlation between $d_L$ and $z$ indicated the expansion of space itself, and cemented the use of dynamical space-time models to describe the Universe at cosmological scales. A century after its discovery the Hubble-Lema\^{\i}tre law (Hereafter referred to as the \sayy{Hubble law}) is widely used in cosmological analysis. 

Since its first formulation, the Hubble law\footnote{In this paper, we use the term \sayy{Hubble law} as meaning the series expansion of luminosity distance or angular diameter distance in redshift.} has been generalised for a number of applications. 
Within the Friedmann-Lema\^{\i}tre-Robertson-Walker (FLRW) class of metrics, the geometrical expansion of the luminosity distance in redshift has been formulated \cite{Visser:2003vq}. 
The luminosity distance Hubble law has been considered for broader classes of geometries, including Lema\^{\i}tre-Tolman-Bondi models \cite{Celerier:1999hp,Tanimoto:2007dq} and certain families of Szekeres cosmological models \cite{Villani:2014zta}.     
Series expansion of the Jacobi map around the vertex of an observer has been considered for arbitrary space-times \cite{Seitz:1994xf} resulting in a general angular diameter distance expansion in the affine parameter along a light beam. 
Power series of cosmological distances in redshift have been obtained for general relativistic space-times \cite{KristianSachs,EllisMacCallum} and expressed using multipole expansions \cite{Clarkson:2011uk,Clarkson:2011br}. The present paper builds on these results.  
Studies of distance--redshift relations in various anisotropic media -- typically under the assumption of an FLRW \sayy{background} metric -- have been made, e.g. \cite{DyerRoeder:1974,Sasaki:1987,FutamaseSasaki:1989,Kantowski:1998,Sugiura:1999,PyneBirkinshaw:2004,Bonvin:2005ps,Fleury:2014gha,Bentivegna:2016fls,SikoraGlod}. Despite of some theoretical interest in the modelling of anisotropies, the isotropic ansatz is used for the vast majority of observational analysis, though see \cite{Wang:2014vqa,Bolejko:2015gmk,Bernal:2016kfw,Colin:2018ghy}. 
While a notion of statistical homogeneity and isotropy is indicated by the convergence of galaxy 2-point correlation function statistics \cite{Hogg,Scrimgeour}, large scale structures of sizes that extend hundreds of Megaparsecs in effective {radii} have been detected \cite{ShapleyAmes,Proust:2005jt,Gott:2003pf,Aihara} along with coherent \sayy{bulk flow velocity} on scales of several hundreds of Megaparsecs \cite{Feindt,Magoulas}. In this paper we develop a model independent framework for taking into account the existence of cosmic structures in the analysis of distance--redshift data.  

As we shall detail, the proportionality law between redshift and distance as it was first hypothesised by Lema\^{\i}tre \cite{Lemaitre:1927} and observed by Slipher and Hubble \cite{Slipher:1917,Hubble:1929} emerges in the small redshift and isotropic limit of the luminosity distance in a general expanding space-time setting. In this paper we derive the expression for the luminosity distance \sayy{Hubble law} as formulated in a general space-time geometry admitting such a series expansion, from which the form of the originally formulated Hubble law derives as a monopole approximation in the $\mathcal{O}(z)$ vicinity of the observer.  
The strength of the final expression for the general anisotropic Hubble law is its simple form given by a finite set of physically interpretable multipole coefficients, providing a new framework for observational analysis.  

In section \ref{sec:dyncongruences} we introduce a general time-like congruence and consider mutual distances defined between observers comoving with the congruence through their emission and absorption of light rays. In section \ref{sec:expdA} we present the general \sayy{Hubble law}: the series expansion of luminosity distance in redshift for general space-times admitting such a series expansion. 
Section \ref{sec:multipole} contains the main result of this paper: we formulate the anisotropic coefficients of the general Hubble law in a multipole representation. 
In section \ref{sec:assumptions} we review the assumptions behind the derived results, and consider the effect of a special relativistic boost of the observer with respect to the frame of the smooth congruence description.   
We conclude in section \ref{sec:conclusion}.

\vspace{5pt}
\noindent
\underbar{Notation and conventions:}
Units are used in which $c=1$. Greek letters $\mu, \nu, \ldots$ label spacetime
indices in a general basis. Einstein notation is used such that repeated indices are summed over.  
The signature of the spacetime metric $g_{\mu \nu}$ is $(- + + +)$ and the connection $\nabla_\mu$ is the Levi-Civita connection. 
Round brackets $(\, )$ containing indices denote symmetrisation in the involved indices and square brackets $[\, ]$ denote anti-symmetrisation. 
Bold notation $\bm T$ for the basis-free representation of tensorial quantities $T_{\nu_1 , \nu_2, .. , \nu_n }^{\qquad \quad \; \mu_1 , \mu_2, .. , \mu_m }$ is used occasionally. 


\section{The cosmological observers and light bundles}
\label{sec:dyncongruences}
We consider a time-like congruence governed by the tangent vector field $\bm{u}$, representing the 4--velocity of reference observers and emitters in the space-time. We further consider an irrotational geodesic congruence of photons with affinely parametrised tangent vector $\bm{k}$ representing the 4--momentum of the light emitted and observed between members of the time-like congruence. 
It will be useful to decompose the photon 4--momentum field $\bm{k}$ in terms of the 4--velocity field $\bm{u}$ as follows 
\bea
\label{kdecomp}
k^\mu = E (u^\mu - e^\mu) \, , \qquad E \equiv - k^\mu u_\mu \, , 
\eea 
where $\bm{e}$ is a vector field orthogonal to $\bm{u}$, and $E$ is the energy of the photon congruence as inferred by observers comoving with $\bm{u}$. The sign convention is such that $\bm{e}$ describes the spatial direction of observation of observers comoving with $\bm{u}$, and we can think of $\bm{e}$ as a radial unit-vector when evaluated at the vertex point of the observer's lightcone. 
We define the projection tensors 
\bea
\label{prroj}
h_{ \mu }^{\; \nu } \equiv u_{ \mu } u^{\nu } + g_{ \mu }^{\; \nu }  \, , \qquad p_{ \mu }^{\; \nu } \equiv u_{ \mu } u^{\nu } - e_{ \mu } e^{\nu } + g_{ \mu }^{\; \nu }  , 
\eea 
onto the 3--dimensional space orthogonal to $\bm{u}$ and the 2--dimensional screen space orthogonal to $\bm{k}$ and $\bm{u}$ respectively. 
The equations describing the deformation of the photon congruences are given by \cite{sac1961,wal1984}
\bea
\label{opticalraychaudhuri}
\frac{ {\rm d} \hth}{{\rm d} \lambda}
& = & -\frac{1}{2}\,\hth^{2} - \hsig_{\mu \nu} \hsig^{\mu \nu} 
- k^{\mu}k^\nu R_{\mu \nu}  \ , \\ 
\label{opticalshear}
\frac{ {\rm d} \hsig_{\mu \nu} }{{\rm d} \lambda}
& = & - \hth \hsig_{\mu \nu}  -  p_{ \mu }^{\, \rho } p_{ \nu }^{\, \sigma } k^\alpha k^\beta  C_{\rho \alpha \sigma \beta}  \ , 
\eea 
where the operation $\frac{{\rm d} }{{\rm d} \lambda} \equiv k^\mu \nabla_\mu$ is the directional derivative along the flow lines of the photon 4-momentum, where $R_{\mu \nu}$ is the Ricci tensor of the space-time and $C_{\mu \nu \rho \sigma}$ is the Weyl tensor, and where the decomposition of the expansion tensor of the null congruence 
\bea
\label{def:optscalin}
\hspace*{-1cm} && \hat{\Theta}_{\mu \nu} \equiv  p_{\nu}^{\, \beta} p_{\mu}^{\, \alpha}  \nabla_{\beta} k_\alpha = \frac{1}{2} \hth p_{\mu \nu } + \hsig_{\mu \nu} \, , \qquad  \hth \equiv \nabla_{\mu}k^{\mu} \, ,  \qquad \hsig_{\mu \nu} \equiv p_{  \mu }^{\, \alpha }p_{  \nu  }^{\, \beta} \nabla_{  \alpha}k_{\beta }  - \frac{1}{2}\hth p_{\mu \nu } \,  
\eea 
has been used. The variables $\hth$ and $\hsig_{\mu \nu}$ describe the expansion and stretching of the congruence area as measured in the 2--dimensional screen space. 
The equations governing the geodesic deviation of the time-like congruence are \cite{wal1984}
\bea
\label{raychaudhuri}
\hspace*{-1cm} \frac{ {\rm d} \theta}{{\rm d} \tau}
& = & -\frac{1}{3}\,\theta^{2} - \sig_{\mu \nu} \sig^{\mu \nu} + \omega_{\mu \nu} \omega^{\mu \nu} 
- u^{\mu}u^\nu R_{\mu \nu}  + D_\mu a^\mu  + a_\mu a^\mu \ , \\ 
\label{shear}
\frac{ {\rm d} \sig_{\mu \nu} }{{\rm d} \tau}
& = & - \frac{2}{3} \theta \sig_{\mu \nu}  - \sig_{\alpha \la \mu} \sig^{\alpha}_{\; \nu \ra}  +  \omega_{\alpha \la \mu} \omega^{\alpha}_{\; \nu \ra}  - u^\rho u^\sigma  C_{\rho \mu \sigma \nu}  -   \frac{1}{2} R_{ \la \mu \nu \ra } \nonumber  \\ 
&+& D_{\la \mu} a_{\nu \ra}  + a_{\la \mu} a_{\nu \ra}  + 2 a^\alpha \sigma_{\alpha ( \nu} u_{\mu)}  \ , \\
\label{rotation}
\frac{ {\rm d} \omega_{\mu \nu} }{{\rm d} \tau}
& = & - \frac{2}{3} \theta \omega_{\mu \nu}  + 2 \sigma^{\alpha}_{\, [\mu} \omega_{\nu ] \alpha}  - D_{[\mu} a_{\nu]}  - 2 a^\alpha \omega_{\alpha [ \mu} u_{\nu]}   \ , 
\eea 
where $\frac{{\rm d} }{{\rm d} \tau} \equiv u^\mu \nabla_\mu$ is the directional derivative along the flow lines of $\bm{u}$, where $a^{\mu} \equiv u^\nu \nabla_\nu u^\mu$ is the 4--acceleration field, where the spatial covariant derivative $D_\mu$ is defined through its acting on a tensor field $T_{\nu_1 , \nu_2, .. , \nu_n }^{\qquad \quad \; \gamma_1 , \gamma_2, .. , \gamma_m }$: 
\bea
\label{D}
\hspace*{-0.2cm} D_{\mu} T_{\nu_1 , \nu_2, .. , \nu_n }^{\qquad \quad \; \gamma_1 , \gamma_2, .. , \gamma_m } \equiv  h_{ \nu_1 }^{\, \alpha_1 } h_{ \nu_2 }^{\, \alpha_2 } .. h_{ \nu_n }^{\, \alpha_n }    \,  h_{ \beta_1 }^{\, \gamma_1 } h_{ \beta_2 }^{\, \gamma_2 } .. h_{ \beta_n }^{\, \gamma_m }    \, h_{ \mu }^{\, \sigma } \nabla_\sigma  T_{\alpha_1 , \alpha_2, .. , \alpha_n }^{\qquad \quad \; \beta_1 , \beta_2, .. , \beta_m } \,  , 
\eea   
and where we have used the decomposition  
\bea
\label{def:expu}
&& \Theta_{\mu \nu} \equiv h_{\nu}^{\, \beta}  h_{\mu}^{\, \alpha} \nabla_{\beta}u_\alpha  = \frac{1}{3}\theta h_{\mu \nu }+\sig_{\mu \nu} + \omega_{\mu \nu}  \ , \nonumber \\ 
&& \theta \equiv \nabla_{\mu}u^{\mu} \, ,  \qquad \sig_{\mu \nu} \equiv \nabla_{ \la \nu }u_{\mu \ra } \, , \qquad  \omega_{\mu \nu} \equiv h_{  \nu  }^{\, \beta}  h_{  \mu }^{\, \alpha }\nabla_{  [ \beta}u_{\alpha ] }   \,  . 
\eea 
The operation $\la \ra$ around indices selects the tracefree symmetric part of the spatial projection, i.e., $A_{\la \mu \nu \ra} \equiv h_{ ( \mu }^{\, \alpha }h_{  \nu ) }^{\, \beta} A_{\alpha \beta} - \frac{1}{3} h_{\mu \nu} h^{\alpha \beta} A_{\alpha \beta} $. The expansion scalar $\theta$ describes the expansion rate of the congruence, the shear term $\sig_{\mu \nu}$ describes its stretching, and the vorticity $\omega_{\mu \nu}$ describes its rotation. The evolution equations (\ref{opticalraychaudhuri})--(\ref{opticalshear}) and (\ref{raychaudhuri})--(\ref{rotation}) are purely geoemetrical and follow from the Ricci identity. 
The evolution of the energy function $E$ along the flow lines of $\bm{k}$ can be formulated in terms of projected dynamical variables of $\bm{u}$ in the following way 
\bea
\label{def:Eevolution}
\frac{ {\rm d} E}{{\rm d} \lambda} = - k^{\mu}\nabla_{\mu} ( k^{\nu} u_\nu ) = - E^2  \Eu \, , \qquad   \Eu \equiv  \frac{1}{3}\theta  - e^\mu a_\mu + e^\mu e^\nu \sigma_{\mu \nu}   \, . 
\eea 
We note that $\Eu$ is a truncated multipole series in the spatial unit vector $\bm{e}$, which will be important for the multipole decomposition of the general Hubble law derived in section \ref{sec:multipole}. 

\vspace{5pt}
\noindent
We now describe cosmological distance measures defined between observers and emitters comoving with $\bm{u}$. 
The angular diameter distance $d_A$ of an astronomical source -- describing the physical size of an object relative to the angular measure it subtends on the sky of a reference observer -- is defined as \cite{ell1971}
\be
d_{A}^{2}  \equiv  \frac{{\delta }A}{{\delta} \Om} \ ,
\ee
where ${\delta}A$ is the proper area of the object perpendicular
to the 4--velocity $\bm{u}$ of the emitting congruence and the 4--momentum $\bm{k}$ of the lightbeam, and ${\delta}\Om$ is the subtended angular element. 
The evolution of the angular diameter distance along the photon congruence 
is given by the expansion rate of the null congruence 
\be
\label{areadistevol}
\frac{{\rm d} d_A }{ {\rm d} \lambda} = \frac{1}{2}\,\hth d_{A} \, , 
\ee 
and the second derivative of $d_A$ can be rewritten using (\ref{opticalraychaudhuri}), yielding the focusing equation \cite{sac1961,Perlick:2010zh}
\be
\label{focuseq}
\frac{{\rm d}^2 d_A }{ {\rm d} \lambda^2} 
= -\left(\hsig^{2} + \frac{1}{2}k^{\mu}k^\nu R_{\mu \nu}  \right)d_{A} \, .
\ee
Assuming the conservation of photons it follows from Etherington's reciprocity theorem \cite{eth1933,ellhve1999} that the luminosity distance between a source and an observer 
\be
\label{luminosityd}
d_L  \equiv  \sqrt{\frac{L}{4\pi F}} \ ,
\ee
is related to the angular diameter distance in the following way 
\be
\label{luminosityd2}
d_L = (1+z)^2 d_A \, , \qquad  1+z \equiv \frac{E}{E_o} \, , 
\ee
where $L$ is the bolometric luminosity of the source, $F$ is the bolometric flux of energy as measured by the observer, $E$ is the photon energy in the frame of the emitter, $E_o$ is the photon energy measured at the vertex point $o$ of the observer's lightcone, and $z$ is the associated redshift.

\section{The luminosity distance Hubble law} 
\label{sec:expdA} 
In this section we formulate the series expansion of luminosity distance in redshift to third order in the general case: we make no restrictions on the observer congruence or the curvature of space-time other than the regularity requirements implied by Taylor's theorem. 
Let $\lambda$ be an affine function along the null rays of the photon congruence satisfying $k^\mu \nabla_\mu \lambda = 1$. 
We expand the angular diameter distance around the point of observation\footnote{The series expansion is defined along the individual null rays, which can be labelled uniquely by the direction vector $\bm{e}_{o}$, and the coefficients of the series expansion have implicit dependence on the direction vector $\bm{e}_{o}$. This dependence reflects the anisotropy of the local space-time neighbourhood relative to the observer at $o$.} $o$ in the following way 
\be
\label{dAexpand}
d_A  = d_{A} \lvert_{o}  +   \frac{{\rm d} d_A }{ {\rm d} \lambda}   \Bigr\rvert_{o}  \Delta \lambda +  \frac{1}{2} \frac{{\rm d}^2 d_A }{ {\rm d} \lambda^2}   \Bigr\rvert_{o}  \Delta \lambda^2 +  \frac{1}{6} \frac{{\rm d}^3 d_A }{ {\rm d} \lambda^3}   \Bigr\rvert_{o}  \Delta \lambda^3  + \mathcal{O}( \Delta \lambda^4) \ ,
\ee 
where $\Delta \lambda \equiv \lambda - \lambda_o$, and where the subscript $o$ denotes evaluation at the vertex point. We can invoke the initial conditions \cite{Seitz:1994xf}
\be
\label{dAinitial}
d_{A} \lvert_{o}  = 0 \, , \qquad   \frac{ \partial  }{ \partial \tau}  \bigr\rvert_{  \bm{k}  } d_A   \Bigr\rvert_{o}  { = \frac{ 1 }{ E_o }   \frac{{\rm d} d_A }{ {\rm d} \lambda}   \Bigr\rvert_{o} }  =  - 1  \, , \qquad  \hsig_{\mu \nu} \rvert_{o} = 0   \ ,
\ee 
where $\frac{ \partial  }{ \partial \mathcal{S} }    \rvert_{  \bm{k}  } \equiv  (1/ k^{\mu} \nabla_\mu \mathcal{S})  \frac{{\rm d}  }{ {\rm d} \lambda}$ denotes the derivative with respect to the scalar function $\mathcal{S}$ along the flow lines of $\bm{k}$. 
The function $\tau$ is the proper time parameter of $\bm{u}$ defined from $u^{\mu}\nabla_\mu \tau = 1$ and $h^\mu_{\, \nu} \nabla_\mu \tau \rvert_{o} = 0$, so that $k^{\mu} \nabla_\mu \tau \rvert_{o}  = E_o$ follows from the decomposition (\ref{kdecomp}). 
The first initial condition of (\ref{dAinitial}) ensures that the area measure tends to zero at the observer, while the second requirement ensures that the area measure reduces to the Euclidian one near a point. 
The third condition states that the shear rate $\hsig_{\mu \nu}$ tends to zero at the focus point of the null rays and results from the Minkowski limiting description around a single space-time point. The divergence of the expansion rate of the photon congruence $\hat{\theta}$ at the observer position $o$ reflects the caustic nature of the vertex point \cite{Seitz:1994xf,Fleury:2015rwa}. 
{From the initial conditions in (\ref{dAinitial}) it follows immediately that the first two coefficients in the series expansion (\ref{dAexpand}) read $d_{A} \lvert_{o} =\! 0$ and $\frac{{\rm d} d_A }{ {\rm d} \lambda}  \rvert_{o} \! = \! - E_o$. 
Further, using (\ref{focuseq}) together with the first and third initial conditions of (\ref{dAinitial}) we have $\frac{{\rm d}^2 d_A }{ {\rm d} \lambda^2}  \rvert_{o} = 0$, assuming that $k^{\mu}k^\nu R_{\mu \nu}$ is non-singular at the vertex point. Computing the directional derivative of both sides of (\ref{focuseq}) along the direction of $\bm{k}$ gives $\frac{{\rm d}^3 d_A }{ {\rm d} \lambda^3}   \rvert_{o} = \frac{1}{2}  E_o k^{\mu}k^\nu R_{\mu \nu}$, assuming that the derivative of $k^{\mu}k^\nu R_{\mu \nu}$ and $\hsig_{\mu \nu}$ are non-singular at the vertex point. 
Thus, the series expansion (\ref{dAexpand}) reduces to }
\be
\label{dAexpand2}
d_A  =  - E_o  \Delta \lambda +  \frac{1}{12}  E_o k^{\mu}k^\nu R_{\mu \nu}  \Bigr\rvert_{o}  \Delta \lambda^3  + \mathcal{O}( \Delta \lambda^4) \ .
\ee 
We can rewrite $\Delta \lambda$ in terms of its expansion in $z$, assuming that the function $\lambda \mapsto z(\lambda)$ is invertible\footnote{{This and other regularity conditions must be satisfied for the below derived expressions to be well defined. See section \ref{sec:assumptions} for a complete discussion of assumptions and their physical significance.}}: 
\be
\label{lambdaexpand}
\Delta \lambda = \frac{ \partial  }{ \partial z } \bigr\rvert_{  \bm{k}  }    \lambda  \Bigr\rvert_{o} z +  \frac{1}{2}  \frac{ \partial^2  }{ \partial z^2 } \bigr\rvert_{  \bm{k}  }    \lambda  \Bigr\rvert_{o} z^2 + \frac{1}{6}  \frac{ \partial^3  }{ \partial z^3 } \bigr\rvert_{  \bm{k}  }    \lambda  \Bigr\rvert_{o} z^3 +  \mathcal{O}( z^4)  \ ,
\ee 
where $z_{o} = 0$ per definition. 
We can find the coefficients of the expansion (\ref{lambdaexpand}) by using (\ref{def:Eevolution}), which gives {
\bea
\label{lambdacoef}
\hspace*{-0.6cm} \frac{ \partial  }{ \partial z } \bigr\rvert_{  \bm{k}  }    \lambda  \Bigr\rvert_{o} &=&  \frac{1  }{ k^\mu \nabla_\mu z }    \Bigr\rvert_{o}  = - \frac{1}{E_o \Eu_o} \, , \nonumber  \\
\hspace*{-1cm} \frac{ \partial^2  }{ \partial z^2 } \bigr\rvert_{  \bm{k}  }    \lambda  \Bigr\rvert_{o} &=&    \frac{\frac{{\rm d}  }{ {\rm d} \lambda} \left(\frac{1  }{ k^\mu \nabla_\mu z }  \right)  }{ k^\mu \nabla_\mu z }     \Biggr\rvert_{o} = - \frac{2}{E_o \Eu_o}  \left( - 1 +  \frac{1}{2} \frac{1}{E_o} \frac{     \frac{ {\rm d} \Eu}{{\rm d} \lambda}    }{\Eu^2}  \Bigr\rvert_{o} \right) ,  \nonumber \\ 
\hspace*{-1cm} \frac{ \partial^3  }{ \partial z^3 } \bigr\rvert_{  \bm{k}  }    \lambda  \Bigr\rvert_{o} &=&   \frac{\frac{{\rm d}  }{ {\rm d} \lambda}  \Bigl( \frac{\frac{{\rm d}  }{ {\rm d} \lambda} \bigl(\frac{1  }{ k^\mu \nabla_\mu z }  \bigr)  }{ k^\mu \nabla_\mu z }  \Bigr)  }{ k^\mu \nabla_\mu z }    \Biggr\rvert_{o}   =   - \frac{6}{E_o \Eu_o}  \left( 1 -  \frac{1}{E_o} \frac{     \frac{ {\rm d} \Eu}{{\rm d} \lambda}    }{\Eu^2}  \Bigr\rvert_{o}  + \frac{1}{2}  \frac{1}{E^2_o} \frac{  \left(    \frac{  {\rm d} \Eu}{{\rm d} \lambda} \right)^2    }{\Eu^4}  \Bigr\rvert_{o}  -  \frac{1}{6}  \frac{1}{E^2_o} \frac{      \frac{  {\rm d^2} \Eu}{{\rm d} \lambda^2}    }{\Eu^3}  \Bigr\rvert_{o}   \right)   , 
\eea  
where it has been used that $ \frac{ \partial  }{ \partial z } \bigr\rvert_{  \bm{k}  } = \frac{1}{k^\mu \nabla_\mu z}  \frac{{\rm d}  }{ {\rm d} \lambda}$ from the definition given below equation (\ref{dAinitial}), and where the definition of $z$ in (\ref{luminosityd2}) has been invoked.} 
We can now use the coefficients in (\ref{lambdacoef}) together with (\ref{dAexpand2}) and (\ref{lambdaexpand}) to obtain a third order expansion of $d_A$ in redshift: 
\bea
\label{dAexpand3} 
\hspace*{-0.5cm} d_A  &=& d_A^{(1)} z   + d_A^{(2)} z^2 +  d_A^{(3)} z^3 + \mathcal{O}( z^4) \ , \nonumber \\ 
\hspace*{-0.5cm}  d_A^{(1)} &\equiv& \frac{1}{\Eu_o} \, , \qquad d_A^{(2)} \equiv   \frac{1}{ \Eu_o}  \left( - 1 +  \frac{1}{2} \frac{1}{E_o} \frac{     \frac{ {\rm d} \Eu}{{\rm d} \lambda}    }{\Eu^2}  \Bigr\rvert_{o} \right) \ , \nonumber \\ 
\hspace*{-0.5cm}  d_A^{(3)} &\equiv&  \frac{1}{ \Eu_o}  \left( \! 1  -  \frac{1}{E_o} \frac{     \frac{ {\rm d} \Eu}{{\rm d} \lambda}    }{\Eu^2}  \Bigr\rvert_{o}  + \frac{1}{2}  \frac{1}{E^2_o} \frac{  \left(    \frac{  {\rm d} \Eu}{{\rm d} \lambda} \right)^2    }{\Eu^4}  \Bigr\rvert_{o}  -  \frac{1}{6}  \frac{1}{E^2_o} \frac{      \frac{  {\rm d^2} \Eu}{{\rm d} \lambda^2}    }{\Eu^3}  \Bigr\rvert_{o}  -  \frac{1}{12} \frac{1}{E^2_o} \frac{k^{\mu}k^\nu R_{\mu \nu} }{\Eu^2}  \Bigr\rvert_{o}   \right)     \,  , 
\eea 
{where we have plugged in (\ref{lambdaexpand}) in (\ref{dAexpand2}) and kept terms up till third order in $z$.}  
{It is convenient to write an analogous taylor series of the luminosity distance in redshift} 
\bea
\label{dLexpand}
d_L  &=& d_L^{(1)} z   + d_L^{(2)} z^2 +  d_L^{(3)} z^3 + \mathcal{O}( z^4) \, , 
\eea 
{where the coefficients can be formulated in terms of the coefficients of the expansion of the angular diameter distance (\ref{dAexpand3}) by invoking the first relation in (\ref{luminosityd2}), and from using the first initial condition of (\ref{dAinitial}). This yields the first order coefficient $d_L^{(1)} \! = \! d_A^{(1)}$, the second order coefficient $d_L^{(2)} \! = \! d_A^{(2)} + 2 d_A^{(1)}$ and the third order coefficient $d_L^{(3)} \! = \!  d_A^{(3)} + 2 d_A^{(2)} + d_A^{(1)}$. Using the expressions (\ref{dAexpand3}) in these expressions for the coefficients, we finally have: } 
\bea 
\label{dLexpandcoef}
d_L^{(1)} &\equiv& \frac{1}{\Eu_o} \, , \qquad d_L^{(2)} \equiv   \frac{1}{ \Eu_o}  \left( 1 +  \frac{1}{2} \frac{1}{E_o} \frac{     \frac{ {\rm d} \Eu}{{\rm d} \lambda}    }{\Eu^2}  \Bigr\rvert_{o} \right) \ , \nonumber \\ 
d_L^{(3)} &\equiv&  \frac{1}{ \Eu_o}  \left(  \frac{1}{2}  \frac{1}{E^2_o} \frac{  \left(    \frac{  {\rm d} \Eu}{{\rm d} \lambda} \right)^2    }{\Eu^4}  \Bigr\rvert_{o}  -  \frac{1}{6}  \frac{1}{E^2_o} \frac{      \frac{  {\rm d^2} \Eu}{{\rm d} \lambda^2}    }{\Eu^3}  \Bigr\rvert_{o}  -  \frac{1}{12} \frac{1}{E^2_o} \frac{k^{\mu}k^\nu R_{\mu \nu} }{\Eu^2}  \Bigr\rvert_{o}   \right) \,  . 
\eea 
{Motivated by the FLRW series expansions of distance measures in redshift \cite{Visser:2003vq}, we identify}
\bea
\label{paramseff}
\hspace*{-0.7cm} \Eu  \, , \qquad \mathfrak{Q}  &\equiv& - 1 - \frac{1}{E} \frac{     \frac{ {\rm d} \Eu}{{\rm d} \lambda}    }{\Eu^2}   \, , \qquad
\mathfrak{R} \equiv  1 +  \mathfrak{Q}  - \frac{1}{2 E^2} \frac{k^{\mu}k^\nu R_{\mu \nu} }{\Eu^2}   \, ,  \qquad \mathfrak{J}  \equiv   \frac{1}{E^2} \frac{      \frac{  {\rm d^2} \Eu}{{\rm d} \lambda^2}    }{\Eu^3}  - 4  \mathfrak{Q}  - 3 \, ,
\eea 
as effective observational \sayy{Hubble}, \sayy{deceleration}, \sayy{curvature}, and \sayy{jerk} parameters respectively. {From (\ref{def:Eevolution}) and (\ref{paramseff}) we see that $1/E$ plays the role of a generalised \sayy{scale-factor} on the null cone, in that its derivatives along $\bm k$ result in the effective observational Hubble, deceleration, and jerk parameters.}\footnote{The reduction to the FLRW Hubble, deceleration, and jerk parameters in the homogeneous and isotropic case where $E$ is proportional to the inverse homogeneous and isotropic scalefactor, can be seen immediately by using the decomposition (\ref{kdecomp}) in (\ref{def:Eevolution}) and (\ref{paramseff}). For instance $\Eu = \frac{ {\rm d} (1/E)}{{\rm d} \lambda} =  E \frac{ {\rm d} (1/E)}{{\rm d} \tau} - E e^\mu \partial_\mu(1/E)$, where the second term vanishes identically in the FLRW limit with observers comoving with the homogeneous and isotropic space-time foliation.} {While $1/E$ is a natural observational generalisation of the FLRW scale-factor for the interpretation of distance--redshift data, it does not inherit all properties which we associate with an FLRW scale-factor. For instance, $1/E$ does in general not describe the scaling of volume in the frame of the observer congruence generated by $\bm u$. Neither is the dynamics of $E$ in general constrained by any FLRW-like field equations.} {From (\ref{def:Eevolution}) it is evident that the effective Hubble parameter includes contributions from non-gravitational forces via the 4--acceleration of the observer congruence. In addition, the effective Hubble parameter includes contributions from the shear tensor of the observer congruence, which is in turn directly linked to Weyl curvature of the space-time through the evolution equation (\ref{shear}). As we shall see in the following analysis, the effective deceleration, curvature, and jerk parameters contain non-trivial contributions from the 4--acceleration as well as kinematic and dynamic variables associated with the observer congruence.} We write the expansion coefficients (\ref{dLexpandcoef}) in terms of these parameters as evaluated at the vertex point: 
\bea
\label{dLexpand2}
d_L^{(1)} &=& \frac{1}{\Eu_o} \, , \qquad d_L^{(2)} =   \frac{1 - \mathfrak{Q}_o }{2 \Eu_o}  \ , \qquad d_L^{(3)} =  \frac{- 1 +  3 \mathfrak{Q}_o^2 + \mathfrak{Q}_o    -  \mathfrak{J}_o   + \mathfrak{R}_o }{ 6  \Eu_o}     \, , 
\eea 
{and see that the effective observational parameters replace the FLRW Hubble $H_o$, deceleration $q_o$, curvature $\Omega_k$, and jerk $j_o$ parameters of the analogous FLRW expansion \cite[eq. (46)]{Visser:2003vq} with: $d_L^{(1)} = \frac{1}{H_o} \, , \; \, d_L^{(2)} =   \frac{1 - q_o }{2 H_o}  \, , \; \, d_L^{(3)} =  \frac{- 1 +  3 q_o^2 + q_o    -  j_o   + \Omega_k }{ 6  H_o}$.} 
The set of parameters $\{ \Eu_o, \mathfrak{Q}_o, \mathfrak{R}_o, \mathfrak{J}_o \}$ is generally \emph{anisotropic} and depend on the direction of observation $\bm{e}_{o}$ in a manner which will be detailed in the following section. 
We note that $\Eu_o$, $\mathfrak{Q}_o$, and $\mathfrak{J}_o$ arise from the reparametrization of the null-lines in terms of the redshift function as incorporated in the coefficients (\ref{lambdacoef}). The only information coming from the focusing of null rays (\ref{focuseq}) is the projected Ricci curvature $k^{\mu}k^\nu R_{\mu \nu} \rvert_{o}$, which is contained in the effective curvature parameter $\mathfrak{R}_o$. 

When evaluated in an FLRW space-time with comoving observers, the variables (\ref{paramseff}) reduce to the FLRW Hubble constant, deceleration parameter, curvature parameter, and jerk respectively, and the coefficients (\ref{dLexpand2}) reduce to those of the well-known FLRW series expansion \cite[eq. (46)]{Visser:2003vq}.
As we shall derive in the following, the effective parameters $\Eu$, $\mathfrak{Q}$, $\mathfrak{R}$, and $\mathfrak{J}$ of the luminosity distance expansion law acquire non-trivial correction terms with respect to the naive isotropized extrapolations\footnote{The function ${}^{(3)}R$ is the adapted 3-curvature on the spatial sections defined by $\bm{u}$ in the case of vanishing vorticity $\omega_{\mu \nu}$ of the congruence.} \sayy{$H$}$:= \theta/3$, \sayy{$q$}$:= -1 -  3 \frac{{\rm d}   \theta }{{\rm d} \tau} / \theta^2$, \sayy{$\Omega_k$}$:=-(3/2){}^{(3)}R/\theta^2$, and \sayy{$j$}$:=1 + 9 ( \frac{{\rm d}^2   \theta }{{\rm d} \tau^2} + \theta \frac{{\rm d}   \theta }{{\rm d} \tau} ) / \theta^3$ of the FLRW Hubble, deceleration, curvature, and jerk parameters respectively. For instance, the effective observational deceleration parameter $\mathfrak{Q}$ differs from the isotropized deceleration parameter \sayy{$q$}$:=-1 -  3 \frac{{\rm d}   \theta }{{\rm d} \tau} / \theta^2$ considered in \cite{Tsagas:2009nh,Tsagas:2011wq} in two ways. Firstly, the full expansion variable $\Eu$ enter the definition in (\ref{paramseff}) instead of the isotropized part $\theta/3$ only (but see \cite{Tsagas:2011wq} where the shear component is also considered). Secondly, the evolution of $\Eu$ is evaluated along the flow lines of $\bm{k}$ instead of the flow lines of $\bm{u}$, adding a derivative along spatial propagation direction $-\bm{e}$ of the incoming light beam via the decomposition (\ref{kdecomp}). This adds potentially crucial multipole terms to the effective observational deceleration parameter $\mathfrak{Q}$.

For reasons that will become clear in the following section, we introduce the modified set of dimensionful parameters $\{ \Eu, \mathfrak{Q}, \mathfrak{R}, \mathfrak{J} \} \mapsto$ 
\bea
\label{paramsefftilde}
 \Eu \, , \qquad \widehat{\mathfrak{Q}} &\equiv&  \Eu^2 (\mathfrak{Q} + 1) \, , \qquad  \widehat{\mathfrak{J}} \equiv  \Eu^3 (\mathfrak{J} - 1 - \mathfrak{R}) \, ,
\eea 
in terms of which the Hubble law reads 
\bea
\label{dLexpandtilde}
d_L^{(1)} &=& \frac{1}{\Eu_o} \, , \qquad d_L^{(2)} =   \frac{2 - \frac{\widehat{\mathfrak{Q}}_o }{\Eu^2_o} }{2 \Eu_o}  \ , \qquad d_L^{(3)} =  \frac{  3 \frac{\widehat{\mathfrak{Q}}_o^2 }{\Eu^4_o}  - 5  \frac{\widehat{\mathfrak{Q}}_o }{\Eu^2_o}   -  \frac{\widehat{\mathfrak{J}}_o  }{\Eu^3_o}    }{ 6  \Eu_o}     \, . 
\eea 
We shall argue that this representation of the luminosity distance Hubble law is particularly natural for the analysis of data in a general anisotropic universe.

\section{Multipole analysis of the luminosity distance Hubble law} 
\label{sec:multipole} 
We shall now consider the effective kinematical parameters (\ref{paramseff}) and analyse their dependence on the direction of observation.  
We have from the definition in (\ref{def:Eevolution}) that $\Eu$ is given by the truncated multipole series 
\bea
\label{Eu}
\Eu (\bm e )&=&  \frac{1}{3}\theta  -  \bm{e} \cdot \bm{a}    +  \bm{e} \bm{e} \cdot  \bm{\sigma}     \, ,
\eea  
where we have used the compact notation $\bm{e} \! \cdot \! \bm{a} =  e^\mu a_\mu$ and $\bm{e} \bm{e} \cdot  \bm{\sigma} = e^\mu e^\nu \sigma_{\mu \nu}$ for later convenience, and we have made the direction dependence of $\Eu$ clear by explicating $\Eu (\bm e )$. The expansion rate $\theta$ constitutes the monopole term, the negative acceleration $-\bm a$ is the dipole coefficient, and $\bm \sigma$ is the quadrupole coefficient, with the eigendirections of the quadrupole moment corresponding to the axes of stretch/contraction of the observer congruence. 
We next consider the effective deceleration parameter $\mathfrak{Q}$ defined in (\ref{paramseff}), which can be written as 
\bea
\label{q}
\mathfrak{Q}(\bm e ) &=&  - 1 -  \frac{ \overset{0}{\mathfrak{q}}   +  \bm{e} \cdot  \bm{{\overset{1}{\mathfrak{q}}}}   +    \bm{e} \bm{e} \cdot  \bm{{\overset{2}{\mathfrak{q}}}}     +    \bm{e} \bm{e} \bm{e} \cdot  \bm{{\overset{3}{\mathfrak{q}}}}    +    \bm{e} \bm{e} \bm{e} \bm{e} \cdot  \bm{{\overset{4}{\mathfrak{q}}}}   }{\Eu^2(\bm e )}    \, , 
\eea 
with 
\bea
\label{qpoles}
&& \overset{0}{\mathfrak{q}} \equiv  \frac{1}{3}   \frac{ {\rm d}  \theta}{{\rm d} \tau} + \frac{1}{3} D_{   \mu} a^{\mu  } - \frac{2}{3}a^{\mu} a_{\mu}    - \frac{2}{5} \sigma_{\mu \nu} \sigma^{\mu \nu}    \, , \nonumber \\ 
&& \overset{1}{\mathfrak{q}}_\mu \equiv   - h^{\nu}_{\mu} \frac{ {\rm d}  a_\nu }{{\rm d} \tau} - \frac{1}{3} D_{\mu} \theta  + a^\nu \omega_{\mu \nu}  +  \frac{9}{5}  a^\nu \sigma_{\mu \nu}    -  \frac{2}{5}   D_{  \nu} \sigma^{\nu }_{\;  \mu  }   \, , \nonumber \\
&& \overset{2}{\mathfrak{q}}_{\mu \nu}  \equiv    h^{\alpha}_{ \mu} h^{\beta}_{\nu} \frac{ {\rm d}  \sigma_{\alpha \beta}   }{{\rm d} \tau} +   D_{  \la \mu} a_{\nu \ra } + a_{\la \mu}a_{\nu \ra }     - 2 \sigma_{\alpha (  \mu} \omega^\alpha_{\; \nu )}   - \frac{6}{7} \sigma_{\alpha \la \mu} \sigma^\alpha_{\; \nu \ra }   \, , \nonumber \\ 
 && \overset{3}{\mathfrak{q}}_{\mu \nu \rho}  \equiv -  D_{ \la \mu} \sigma_{\nu   \rho \ra }    -  3  a_{ \la \mu} \sigma_{\nu \rho \ra }    \, , \qquad \quad       \overset{4}{\mathfrak{q}}_{\mu \nu \rho \kappa}  \equiv  2   \sigma_{\la \mu \nu } \sigma_{\rho \kappa \ra} \, , 
\eea 
where we have used the compact notation $\bm{e} \cdot  \bm{{\overset{1}{\mathfrak{q}}}} \equiv e^\mu  \overset{1}{\mathfrak{q}}_\mu$, $\bm{e} \bm{e}  \cdot  \bm{{\overset{2}{\mathfrak{q}}}} \equiv e^\mu e^\nu  \overset{2}{\mathfrak{q}}_{\mu \nu}$, etc., and made use of the identity 
\bea
\label{kderive}
 \frac{{\rm d} e^\mu }{ {\rm d} \lambda} = E (e^\mu - u^\mu) \Eu - E e^\nu \left(\frac{1}{3} \theta h^{\mu}_{\; \nu} + \sigma^{ \mu}_{\; \nu} + \omega^{\mu}_{\; \nu} \right) + E a^\mu \,  .
\eea  
The system (\ref{qpoles}) has been written in its \sayy{traceless decomposition}, and the notation $T_{\la \mu_1 , \mu_2, .. , \mu_n \ra}$ denotes the symmetric and traceless part of the 3--dimensional tensor \\
$T_{\mu_1 , \mu_2, .. , \mu_n} =  h_{ \mu_1 }^{\, \alpha_1 } h_{ \mu_2 }^{\, \alpha_2 } .. h_{ \mu_n }^{\, \alpha_n }     T_{\alpha_1 , \alpha_2, .. , \alpha_n}$ as detailed in appendix \ref{tracereduction}.

The effective observational deceleration parameter $\mathfrak{Q}$ needs not obey the Friedmannian inequality $\mathfrak{Q} \geq 0$ in a general relativistic space-time obeying the strong energy condition $R_{\mu \nu} u^\mu u^\nu  \geq 0$: both monopole and higher order multipole terms can violate the inequality. 
Regional Universe properties alter the Friedmannian deceleration parameter with several non-trivial correction terms that need not be small. For instance, the relative spatial variation in the expansion rate $\theta$ is of order unity between void regions and regions dominated by gravitationally bound structures, and thus $D_\mu \theta$ could realistically contribute to a significant dipole in the effective deceleration parameter. 
Neglecting all anisotropic contributions in (\ref{Eu}) and (\ref{q}), we have the monopole limit $\mathfrak{Q} \rightarrow - 1 -  9 \overset{0}{\mathfrak{q}}/\theta^2$. The effective deceleration parameter $\mathfrak{Q}$ does not in general reduce to the isotropized deceleration of length scales in the observer frame \sayy{$q$}$:= -1 -  3 \frac{{\rm d}   \theta }{{\rm d} \tau} / \theta^2$ due to the presence of terms in $\overset{0}{\mathfrak{q}}$ involving the shear, the 4--acceleration, and the gradient of the 4--acceleration of the observer congruence. Thus, even in its monopole limit, the deceleration parameter does not directly measure volume deceleration in the general case. Furthermore, local values of $\theta$ and $\frac{{\rm d}   \theta }{{\rm d} \tau}$ will in general differ from the corresponding parameters of a given hypothesised FLRW background model. {This effect can in itself give rise to locally positive values of volume acceleration as pointed out in, e.g., \cite{Tsagas:2009nh,Tsagas:2011wq}}. 
For a \sayy{tilted} observer congruence in an FLRW universe model, the leading order correction of the isotropized deceleration of length scales \sayy{$q$}$:= -1 -  3 \frac{{\rm d}   \theta }{{\rm d} \tau} / \theta^2$ relative to that in the \sayy{CMB frame} comes from the term $D_\mu a^\mu$ in Raychaudhuri's equation (\ref{raychaudhuri}) \cite{Tsagas:2009nh,Tsagas:2015mua}. It is in this context interesting to note that $\overset{0}{\mathfrak{q}}$ contains a term $D_\mu a^\mu/3$ in addition to that appearing in the contribution from $\frac{{\rm d}   \theta }{{\rm d} \tau}$ through the Raychaudhuri equation. Taking this extra term into account would lead to an \emph{enhancement} of the predicted apparent monopolar volume acceleration effects quoted in table 1 of \cite{Tsagas:2015mua} for various survey length scales. 

{A dipolar structure in the observed deceleration parameter has been hypothesized in \cite{Tsagas:2009nh,Tsagas:2011wq}, based on FLRW studies with peculiar velocity flow. The above expression for $\mathfrak{Q}$ (\ref{q}) confirms the general expectation of a dipolar structure in the effective deceleration parameter of the luminosity distance Hubble law -- along with higher order anisotropic terms up till quadrupolar order -- even though the deceleration parameter and its dipole takes a different form than that proposed in \cite{Tsagas:2009nh,Tsagas:2011wq}. 
The reason for these differences -- apart from the present analysis being more general than a perturbative framework around an FLRW background space-time -- is that the expressions used for the deceleration parameter in \cite{Tsagas:2009nh,Tsagas:2011wq} relies on the extrapolation of the FLRW deceleration parameter to the effective deceleration parameter \sayy{$q$}$:= -1 -  3 \frac{{\rm d}   \theta }{{\rm d} \tau} / \theta^2$. While such an extrapolation has physical meaning as the isotropized deceleration of length scales in the fluid frame, the natural extrapolation of the FLRW deceleration parameter when considering distance--redshift data is the parameter $\mathfrak{Q}$, which replaces the FLRW deceleration parameter in the general series expansion of luminosity distance in redshift.\footnote{The discrepancy between $\mathfrak{Q}$ and \sayy{$q$} is due to the non-trivial mapping of the relative expansion and acceleration of distances between time-like world lines by the detected null rays: The inverse photon energy $1/E$ defines an observationally relevant length scale, which $\Eu$ and $\mathfrak{Q}$ measure the expansion and deceleration of along the direction of the photon 4-momentum (see the definitions in (\ref{def:Eevolution}) and (\ref{paramseff})). The parameter \sayy{$q$}$:= -1 -  3 \frac{{\rm d}   \theta }{{\rm d} \tau} / \theta^2$ measures the deceleration of length scales in the observer frame, where the length scale is defined from the cube root of the Riemannian volume element in the frame of the observer congruence, which is not directly probed in the distance--redshift relation.} 
Similar conclusions were reached in the investigations \cite{Clarkson:2011uk,Clarkson:2011br} considering the general formula for angular diameter distance. 
Thus, for the purpose of analysing cosmological surveys of standardisable objects, $\mathfrak{Q}$ is the relevant deceleration parameter to consider. }

We may decompose the parameters $\mathfrak{R}$ and $\mathfrak{J}$ into normalised multipole series in a similar way as for $\mathfrak{Q}$, as done in appendix \ref{app:multipole}. Using these results, we can finally write the set of dimensionful parameters (\ref{paramsefftilde}) in the following way: 
\bea 
\label{HM}
\Eu (\bm e ) &=&   \frac{1}{3}\theta  - \bm{e} \cdot \bm{a} + \bm{e} \bm{e} \cdot  \bm{\sigma}   \, , \\ 
\label{QM}
\widehat{\mathfrak{Q}} (\bm e ) &=&   - \left( \overset{0}{\mathfrak{q}}   +  \bm{e} \cdot  \bm{{\overset{1}{\mathfrak{q}}}}   +    \bm{e} \bm{e} \cdot  \bm{{\overset{2}{\mathfrak{q}}}}     +    \bm{e} \bm{e} \bm{e} \cdot  \bm{{\overset{3}{\mathfrak{q}}}}    +    \bm{e} \bm{e} \bm{e} \bm{e} \cdot  \bm{{\overset{4}{\mathfrak{q}}}} \right)  \, ,  \\ 
\label{JM}
\widehat{\mathfrak{J}} (\bm e )  &=&   \overset{0}{\mathfrak{t}}   +  \bm{e} \! \cdot \! \bm{{\overset{1}{\mathfrak{t}}}}   +    \bm{e} \bm{e}   \! \cdot \!  \bm{{\overset{2}{\mathfrak{t}}}}     +    \bm{e} \bm{e} \bm{e} \! \cdot \!  \bm{{\overset{3}{\mathfrak{t}}}}    +    \bm{e} \bm{e} \bm{e} \bm{e}   \! \cdot \!  \bm{{\overset{4}{\mathfrak{t}}}}      +   \bm{e} \bm{e} \bm{e} \bm{e} \bm{e}   \! \cdot \!  \bm{{\overset{5}{\mathfrak{t}}}}     +   \bm{e} \bm{e} \bm{e} \bm{e} \bm{e} \bm{e}   \! \cdot \!  \bm{{\overset{6}{\mathfrak{t}}}}  \, ,   \, 
\eea  
where the set of coefficients $\{ \overset{0}{\mathfrak{t}},  \bm{{\overset{1}{\mathfrak{t}}}}  ,  \bm{{\overset{2}{\mathfrak{t}}}}  , \bm{{\overset{3}{\mathfrak{t}}}}  , \bm{{\overset{4}{\mathfrak{t}}}} , \bm{{\overset{5}{\mathfrak{t}}}} , \bm{{\overset{6}{\mathfrak{t}}}} \}$ is given in appendix \ref{app:multipole}. The multipole coefficients of (\ref{HM})--(\ref{JM}) are all given in terms of dynamical variables associated with the 4--velocity $\bm{u}$ of the observer congruence and curvature degrees of freedom as projected relative to the frame of $\bm{u}$. 
The dependence on the photon congruence enter solely through the projection of the multipole coefficients onto the spatial direction vector $\bm{e}$ of observation. 
Due to their simple form in terms of truncated multipole series, we argue that the set $\{ \Eu, \widehat{\mathfrak{Q}},  \widehat{\mathfrak{J}} \}$ constitutes a \emph{preferred set of parameters for observational analysis}. 

Equation (\ref{dLexpandtilde}) in combination with (\ref{HM})--(\ref{JM}) is the main result of this paper. 
In the trace-free decompositions of the series (\ref{HM})--(\ref{JM}), each $2^\ell$-pole represents $2\ell + 1$ degrees of freedom.  
Thus $\Eu_o$ represents 9 degrees of freedom: $\{\theta_o, \bm{a}_o , \bm{\sigma}_o \}$; $\widehat{\mathfrak{Q}}_o$ represents 16 independent degrees of freedom\footnote{The 16-pole moment of $\widehat{\mathfrak{Q}}_o$ and the 64-pole moment of $\widehat{\mathfrak{J}}_o$ are given in terms of $\bm{\sigma}_o$.}: $\{ \overset{0}{\mathfrak{q}}_o,  \bm{{\overset{1}{\mathfrak{q}}}}_o  ,  \bm{{\overset{2}{\mathfrak{q}}}}_o  , \bm{{\overset{3}{\mathfrak{q}}}}_o \}$; and $\widehat{\mathfrak{J}}_o$ represents 36 independent degrees of freedom: $\{ \overset{0}{\mathfrak{t}}_o,  \bm{{\overset{1}{\mathfrak{t}}}}_o  ,  \bm{{\overset{2}{\mathfrak{t}}}}_o  , \bm{{\overset{3}{\mathfrak{t}}}}_o  , \bm{{\overset{4}{\mathfrak{t}}}}_o , \bm{{\overset{5}{\mathfrak{t}}}}_o \}$. 
Thus, the \emph{exact} anisotropic Hubble law is given by 9, 25, 61 constant degrees of freedom in the $\mathcal{O}(z)$, $\mathcal{O}(z^2)$, $\mathcal{O}(z^3)$ vicinity of the observer respectively.

\vspace{5pt}
\noindent
\underbar{Note on application to data analysis:} 
The general Hubble law (\ref{dLexpandtilde}) with (\ref{HM})--(\ref{JM}) can be used directly in the analysis of distance--redshift data in the small to intermediate redshift regime. To generalise analyses conventionally carried out using the FLRW ansatz for the relation between distances and redshift, the FLRW luminosity distance Hubble law \cite[eq. (46)]{Visser:2003vq} can simply be replaced by the general Hubble law given by (\ref{dLexpandtilde}) and (\ref{HM})--(\ref{JM}). 
The coefficients $\{\theta_o, \bm{a}_o , \bm{\sigma}_o \}$, $\{ \overset{0}{\mathfrak{q}}_o,  \bm{{\overset{1}{\mathfrak{q}}}}_o  ,  \bm{{\overset{2}{\mathfrak{q}}}}_o  , \bm{{\overset{3}{\mathfrak{q}}}}_o \}$, $\{ \overset{0}{\mathfrak{t}}_o,  \bm{{\overset{1}{\mathfrak{t}}}}_o  ,  \bm{{\overset{2}{\mathfrak{t}}}}_o  , \bm{{\overset{3}{\mathfrak{t}}}}_o  , \bm{{\overset{4}{\mathfrak{t}}}}_o , \bm{{\overset{5}{\mathfrak{t}}}}_o \}$ constitutes the set of free parameters of a full model independent analysis.\footnote{Such an analysis requires that the observer is comoving with the cosmological congruence description or that the special relativistic velocity of the observer with respect to the congruence 4--velocity $\bm u$ is known. See section \ref{sec:assumptions} for the derivation of a modified fitting function where the unknown peculiar velocity of the observer with respect to the congruence 4--velocity is incorporated as three additional parameters in a \emph{boosted} version of the Hubble law.} 
When the number of data points and/or sky coverage is too sparse for estimation of the full set of 61 degrees of freedom, a truncation of the series as motivated by the available data may be carried out. Physically motivated assumptions may be imposed on the system of coefficients through their expressions in terms of expansion variables of the observer congruence and curvature degrees of freedom of the space-time.

\section{Discussion of assumptions and accounting for peculiar motion} 
\label{sec:assumptions}
We have presented a model independent expression for the luminosity distance Hubble law to third order in redshift assuming (i) the presence of a time-like congruence constituting the observers and emitters in the space-time; (ii) the presence of an irrotational null geodesic congruence describing the light emitted and observed, and the conservation of photon number; (iii) invertibility of the redshift function $\lambda \mapsto z(\lambda)$ along the flow lines of $\bm{k}$ where $\lambda$ is an affine parameter satisfying $k^\mu \nabla_\mu \lambda = 1$; 
(iv) 3-times differentiability of $z$ along $\bm{k}$ at the vertex point $o$; 
(v) 3-times differentiability of the angular diameter distance $d_A$ along $\bm{k}$ at the vertex point $o$.  

The condition (iii) requires some attention. Invertibility of $\lambda \mapsto z(\lambda)$ is satisfied if and only if $\Eu$ is a continuous function satisfying $\Eu \neq 0$ along each null ray. 
For cosmological large scale solutions, $\Eu$ is reasonably assumed to be of constant (and positive) sign. However, taking into account small scales where collapse and strong deformation of structure occur, $\Eu$ can change sign regionally. 
The break down of the series expansion (\ref{lambdaexpand}) for changing sign of $\Eu$ can be understood physically as follows: when both expansion and contraction happen along the propagation direction of the photon congruence, the energy function $E$ will alternately decrease and increase along the photon congruence. This in turn causes the redshift function relative to the central observer to alternately decrease and increase, along the light ray causing the function $z \mapsto  \lambda(z)$ to be multivalued. Thus redshift cannot be identified as a parameter along the null ray, and the function $z \mapsto  d_A(z)$ is multivalued, i.e., the cosmological notion of distance as a well behaved function of redshift breaks down. 
One may circumvent the issue of multivaluedness either by considering a small enough space-time region around the observer where $\Eu$ is of constant sign, or by coarse-graining over scales large enough such that the sign of $\Eu$ can safely be taken to be constant. 
Physically, the latter strategy requires coarse-graining over scales exceeding the largest collapsing regions, which in the present epoch Universe are of characteristic size $\sim 10$ Megaparsecs \cite{Reisenegger:2000vc,Pearson:2014hja,Einasto:2016rhe}.  
Condition (iv) requires the dynamical variables associated with $\bm{u}$ and their derivatives to be non-singular at the vertex point $o$, and the condition (v) requires $k^\mu k^\nu R_{\mu \nu}$ to be finite at the vertex point. 
Note that the conditions (i)--(v) do not guarantee the remainder term of the Hubble law as formulated for a given order in $z$ to be small. The level of accuracy of the truncated luminosity distance series expansion in redshift must be assessed on a case by case basis. 

\vspace{5pt}
\noindent
\underbar{Note on peculiar velocities:}
A difficulty of any cosmological analysis is to reconcile the cosmological congruence picture with the observer residing inside a single cosmological \sayy{particle}, which may have special relativistic movement with respect to the smooth congruence 4--velocity field $\bm u$. Knowledge of any such \sayy{peculiar velocity} of the observer and emitters is in principle required in order to satisfy the condition (i). 
In the following we incorporate a small peculiar motion of the observer relative to the congruence 4--velocity field\footnote{The peculiar motion of the observer is usually calibrated against the dipole of the cosmic microwave background (CMB).  
However, a recent investigation of the angular distribution of the CatWISE quasar sample concludes that the exclusively kinematic interpretation of the CMB dipole is rejected with a p-value of $5\times 10^{-7}$ \cite{Secrest:2020has}. See also \cite{Bolejko:2015gmk} for an investigation of cosmic rest frames and Hubble flow anisotropies in the COMPOSITE sample of galaxies and clusters. 
} in the Hubble law. We assume that emitters are statistically well described as comoving with the congruence four velocity $\bm u$.  

Let us consider the luminosity distance expansion law as viewed from a \sayy{tilted} observer with a special relativistic velocity $\abs{\bm{v}}  \ll 1$ relative to the congruence four velocity field $\bm{u}$ at the vertex point $o$, such that the four velocity of the tilted observer takes the form $\bm{\tilde{u}}_o = \bm u_o + \bm v$ to first order in $\bm v$. 
It follows that the energy and redshift measured by the tilted observer is 
\bea 
\label{energytilt}
&& \tilde{E}_o \equiv - \tilde{u}^\mu k_\mu \rvert_{o} = E_o (1 + e_o^\mu v_{\mu})   \, , \qquad 1+\tilde{z} \equiv \frac{E}{\tilde{E}_o } = \frac{E_o }{\tilde{E}_o } (1+z) \, , 
\eea  
and that the spatial direction\footnote{The spatial direction of observation $\bm{\tilde{e}}_o$ is defined through the decomposition $\bm k_o = \tilde{E}_o(\bm{\tilde{u}}_o - \bm{\tilde{e}}_o)$ analogously to the definition (\ref{kdecomp}) of $\bm e$.} $\bm{\tilde{e}}_o$ of the incoming light beam as inferred by the tilted observer is given by the relation  
\bea 
\label{etilde}
&& e_o^\mu = (1+ e_o^\nu v_\nu ) \tilde{e}_o^\mu - v^\mu - e_o^\nu v_\nu u_o^\mu \, . 
\eea  
The angular diameter and luminosity distance as measured in the tilted frame are given by \cite{Korzynski:2019oal}
\bea 
\label{dAtilt}
&&\tilde{d}_A= \frac{\tilde{E}_o}{E_o} d_A  \, ,  \qquad \tilde{d}_L  \equiv  (1+\tilde{z})^2  \tilde{d}_A = \frac{E_o}{\tilde{E}_o} d_L  \, , 
\eea  
due to the light aberration effect and the redshifts relation (\ref{energytilt}). 
Invoking the expansion of luminosity distance (\ref{dLexpand}) and using the function $\tilde{z} \mapsto z(\tilde{z})$ given by (\ref{energytilt}) we have for $\bm v \cdot \bm{\tilde{e}}_o \ll z$ and $(\bm v \cdot \bm{\tilde{e}}_o)^2 \ll z^3$: 
\bea 
\label{dLtildexpand}
&&\tilde{d}_L = \tilde{d}_L^{(0)} +  \tilde{d}_L^{(1)} \tilde{z}   + \tilde{d}_L^{(2)} \tilde{z}^2 +  \tilde{d}_L^{(3)} \tilde{z}^3 + \mathcal{O}( \tilde{z}^4) \ , 
\eea  
to first order in $\bm v \cdot \bm{\tilde{e}}_o$, with 
\bea
\hspace*{-1cm} && \tilde{d}_L^{(0)} \! \equiv \! \tilde{e}_o^\mu v_\mu d_L^{(1)}   \, ; \quad \tilde{d}_L^{(1)} \! \equiv \! d_L^{(1)}  \! + \! 2 \tilde{e}_o^\mu v_\mu d_L^{(2)} \, ; \quad \tilde{d}_L^{(2)} \! \equiv \! d_L^{(2)}  \! + \! \tilde{e}_o^\mu v_\mu ( d_L^{(2)} \! + \! 3 d_L^{(3)} ) \, ; \quad \tilde{d}_L^{(3)} \! \equiv \! d_L^{(3)}  \, , 
\eea  
where the third order term does not acquire correction terms for an expansion truncated at $\mathcal{O}( \tilde{z}^3)$, since $\bm v \! \cdot \! \bm{\tilde{e}}_o  \, \tilde{z}^3 \! \ll \! \tilde{z}^4$. 
Using the form (\ref{dLexpandtilde}) of the coefficients $d_L^{(1)}$, $d_L^{(2)}$ and $d_L^{(3)}$ in terms of the effective parameters $\{\Eu , \widehat{\mathfrak{Q}} ,  \widehat{\mathfrak{J}} \}$ -- keeping terms consistently to first order in $\bm{v}$ for the coefficients $d_L^{(0)}$, $d_L^{(1)}$ and $d_L^{(2)}$ and to zeroth order in $\bm{v}$ for $d_L^{(3)}$ -- we have 
\bea 
\label{dLtildecoef}
\hspace*{-0.6cm} \tilde{d}_L^{(0)} &=& \frac{ \tilde{e}_o^\mu v_\mu}{\Eu_o (\bm{ \tilde{e}})}   \, , \qquad \quad \tilde{d}_L^{(1)} = \frac{1}{\Eu_o (\bm{ \tilde{e}}) }   -  \frac{\delta \Eu_o (\bm{ \tilde{e}}) }{\Eu^2_o(\bm{ \tilde{e}})  } +  \frac{\tilde{e}_o^\mu v_\mu}{\Eu_o(\bm{ \tilde{e}}) } \left( 2  - \frac{ \widehat{\mathfrak{Q}}(\bm{ \tilde{e}}) }{\Eu^2_o(\bm{ \tilde{e}}) } \right)    \, ,  \nonumber \\   
\hspace*{-1cm} \tilde{d}_L^{(2)} &=&  \frac{2  -   \frac{\widehat{\mathfrak{Q}}_o(\bm{ \tilde{e}}) }{\Eu^2_o(\bm{ \tilde{e}})} }{2 \Eu_o(\bm{ \tilde{e}})}       -  \frac{\delta \Eu_o (\bm{ \tilde{e}}) }{\Eu^2_o(\bm{ \tilde{e}})  }   -  \frac{1}{2}  \frac{\widehat{\mathfrak{Q}}_o(\bm{ \tilde{e}}) }{\Eu^3_o(\bm{ \tilde{e}})}   \left(    \frac{\delta \mathfrak{Q}_o (\bm{ \tilde{e}}) }{\mathfrak{Q}_o(\bm{ \tilde{e}})  }    -    3 \frac{\delta \Eu_o (\bm{ \tilde{e}}) }{\Eu_o(\bm{ \tilde{e}})  }     \right)  \nonumber \\ 
\quad &&+ \, \frac{1}{2} \frac{\tilde{e}_o^\mu v_\mu }{\Eu_o (\bm{ \tilde{e}})} \left( 2 + 3 \frac{ \widehat{\mathfrak{Q}}^2_o(\bm{ \tilde{e}})  }{ \Eu^4_o (\bm{ \tilde{e}})} - 6 \frac{ \widehat{\mathfrak{Q}}_o(\bm{ \tilde{e}})  }{\Eu^2_o (\bm{ \tilde{e}})}  -  \frac{ \widehat{\mathfrak{J}}_o(\bm{ \tilde{e}})  }{ \Eu^3_o (\bm{ \tilde{e}})}  \right) \, , \nonumber \\ 
\hspace*{-1cm} \tilde{d}_L^{(3)} &=&   \frac{  3 \frac{\widehat{\mathfrak{Q}}_o^2(\bm{ \tilde{e}}) }{\Eu^4_o(\bm{ \tilde{e}})}  - 5  \frac{\widehat{\mathfrak{Q}}_o(\bm{ \tilde{e}}) }{\Eu^2_o(\bm{ \tilde{e}})}   -  \frac{\widehat{\mathfrak{J}}_o (\bm{ \tilde{e}}) }{\Eu^3_o(\bm{ \tilde{e}})}    }{ 6  \Eu_o(\bm{ \tilde{e}})}   \, , 
\eea  
with $\delta \Eu_o(\bm{ \tilde{e}})$ and $\delta \widehat{\mathfrak{Q}}_o(\bm{ \tilde{e}})$ given by the truncated multipole expansions 
\bea 
\label{deltaEuQ}
\delta \Eu(\bm{ \tilde{e}}) &\equiv& v^\mu a_\mu - 2\tilde{e}^\mu v_\nu \sigma_{\mu}^{\, \nu} - \tilde{e}^\mu \tilde{e}^\nu v_\mu a_\nu  + 2  \tilde{e}^\mu \tilde{e}^\nu \tilde{e}^\rho v_\mu \sigma_{\nu \rho}      \, , \nonumber \\  
\delta \widehat{\mathfrak{Q}}(\bm{ \tilde{e}}) &\equiv&  v^\mu \overset{1}{\mathfrak{q}}_\mu + 2 \tilde{e}^\mu v^\nu \overset{2}{\mathfrak{q}}_{\mu \nu} + 3 \tilde{e}^\mu \tilde{e}^\nu \left( v^\rho \overset{3}{\mathfrak{q}}_{\mu \nu \rho}  - v_\mu \overset{1}{\mathfrak{q}}_{\nu}  \right)  + \tilde{e}^\mu \tilde{e}^\nu \tilde{e}^\rho \left( v^\kappa \overset{4}{\mathfrak{q}}_{\mu \nu \rho \kappa} - 2 v_{\mu} \overset{2}{\mathfrak{q}}_{\nu \rho}  \right) \nonumber \\  
&-& 3 \tilde{e}^\mu \tilde{e}^\nu \tilde{e}^\rho \tilde{e}^\kappa  v_{\mu} \overset{3}{\mathfrak{q}}_{\nu \rho \kappa}  - 4 \tilde{e}^\mu \tilde{e}^\nu \tilde{e}^\rho \tilde{e}^\kappa \tilde{e}^\gamma  v_{\mu} \overset{4}{\mathfrak{q}}_{\nu \rho \kappa \gamma} \,  
\eea  
and with $\Eu_o(\bm{ \tilde{e}})$, $\widehat{\mathfrak{Q}}_o(\bm{ \tilde{e}})$, and $\widehat{\mathfrak{J}}_o(\bm{ \tilde{e}})$ given by the expansions (\ref{HM})--(\ref{JM}) with the replacement $\bm e \mapsto \bm{ \tilde{e}}$. 
Note that the expansions $\delta \Eu_o(\bm{ \tilde{e}})$ and $\delta \widehat{\mathfrak{Q}}_o(\bm{ \tilde{e}})$ are given entirely in terms of the multipole coefficients of (\ref{HM}) and (\ref{QM}) and the peculiar velocity vector $\bm v$. 
We see from (\ref{dLtildecoef}) that the luminosity distance expansion in redshift acquires non-trivial correction terms for a non-zero boost velocity $\bm v$. 
In particular the zeroth order term $\tilde{d}_L^{(0)}$ is added to the series expansion on account of the discontinuity in the energy function introduced at the vertex from the peculiar boost away from the smooth congruence description, resulting in the dipolar offset $\tilde{z}_o = E_o/\tilde{E}_o - 1 = - \tilde{e}_o^\mu v_{\mu}$ to first order in $\bm v$. 
Using (\ref{dLtildexpand}), (\ref{dLtildecoef}) instead of (\ref{dLexpand}), (\ref{dLexpandtilde}) allows us to incorporate the fit of a small unknown peculiar velocity of the observer relative to the congruence 4--velocity \emph{without reference to any hypothesised background space-time solution}. Three additional degrees of freedom corresponding to the peculiar velocity vector $\bm v$ are in this case added to the general Hubble law, such that the exact anisotropic \emph{boosted} Hubble law is given by 12, 28, 64 degrees of freedom in the $\mathcal{O}(z)$, $\mathcal{O}(z^2)$, $\mathcal{O}(z^3)$ vicinity of the observer respectively.

\section{Conclusion} 
\label{sec:conclusion} 
We have presented a derivation of the luminosity distance series expansion to third order in redshift without making assumptions on the metric of space-time or the field equations governing it. 
The main result of this paper consists of the equations (\ref{dLexpandtilde}) and (\ref{HM})--(\ref{JM}) where $\Eu_o$, $\widehat{\mathfrak{Q}}_o$, and $\widehat{\mathfrak{J}}_o$ represent the effective Hubble parameter, the effective (transformed) deceleration parameter, and the effective (transformed) \sayy{jerk} parameter of the Hubble law as mapped to a Friedmannian form.    
The parameters $\{ \Eu_o, \widehat{\mathfrak{Q}}_o, \widehat{\mathfrak{J}}_o \}$ are given by \emph{finite} multipole-series where each multipole is formulated in terms of dynamical variables of the observer congruence $\bm{u}$ and curvature degrees of freedom as projected relative to $\bm{u}$. 
Thus, we find this representation of the luminosity distance Hubble law particularly useful.

The derived general Hubble law opens the door to doing consistent multipole analysis of distance--redshift data. 
By a counting of degrees of freedom, the Hubble law is given by 9, 25, 61 degrees of freedom in the $\mathcal{O}(z)$, $\mathcal{O}(z^2)$, $\mathcal{O}(z^3)$ vicinity of the observer respectively. 
If including an unknown special relativistic peculiar velocity of the observer\footnote{The peculiar velocity field is defined with respect to the general congruence 4--velocity $\bm u$. Thus, contrary to usual definitions of peculiar velocities, no background space-time model with hypothesised symmetry properties is invoked as a reference.} relative to the 4--velocity field $\bm u$ of the smooth congruence description, the Hubble law is given by 12, 28, 64 degrees of freedom in the observer's $\mathcal{O}(z)$, $\mathcal{O}(z^2)$, $\mathcal{O}(z^3)$ neighbourhood. 
This is a large number of parameters, but not overwhelmingly so. With upcoming datasets of supernovae of type 1a in the 2020's counting hundreds of thousands of supernovae to be detected at redshifts $z\lesssim2$ over a large proportion of the sky -- see \cite{Scolnic:2019apa} and references therein -- the constraining power for unraveling the relation between luminosity distance and redshift will rise significantly. 
The ability to measure non-zero multipole components depend on the error budget of the cosmological catalogue used as well as the amplitudes of the multipole coefficients. Since the multipole terms appearing in the $\mathcal{O}(z^3)$ Hubble law derive from first, second, and third derivatives of the energy function as well as projections of the Ricci curvature tensor, it is expected that some of the multipole corrections to the monopole limit Hubble law are significantly larger -- perhaps orders of magnitudes larger -- in amplitude than the multipoles of the same order in the expansion of the temperature field of the cosmic microwave background (CMB). 
For a fit to data involving fewer degrees of freedom than the exact Hubble law, the parameters $\{ \Eu_o, \widehat{\mathfrak{Q}}_o, \widehat{\mathfrak{J}}_o \}$ may be truncated in their series expansion as motivated by the available data. The general Hubble law can also be used within a given model setup placing assumptions on the dynamics of $\bm{u}$ and the curvature of space-time.

{One of the main findings in the present paper is that the effective observational deceleration parameter $\mathfrak{Q}_o$, replacing the FLRW deceleration parameter in the generalised distance--redshift series expansion, can be negative without the need of a cosmological constant or an energy source violating the strong energy condition. This can lead to an apparent acceleration effect if observers in an anisotropic and inhomogeneous space-time (or non-ideal observers in an FLRW universe model) are interpreting their data as if they were ideal observers in an FLRW space-time.  
The possibility of (apparant) acceleration of length scales without violation of the strong energy condition has been suggested in previous studies. For instance, investigations of cosmological backreaction \cite{buchert:grgdust,buchert:grgfluid} suggest that structures on small and intermediate scales can contribute to an \emph{emergent} global notion of cosmic acceleration while space is locally decelerating everywhere \cite{Rasanen:2003fy,Rasanen:2006kp}. Furthermore, the possibility of local acceleration of length scales has been suggested in \cite{Tsagas:2009nh,Tsagas:2011wq}. 
While these previous studies clearly present conditions under which a notion of volume acceleration can be physically obtained, the operational definitions of \sayy{deceleration parameters} in these investigations are not directly motivated from the analysis of cosmological datasets. 
In this paper we formulate an observationally relevant extension of the FLRW deceleration parameter to general space-times (together with analogous extensions of the FLRW Hubble, jerk, and curvature parameters), which replaces the FLRW deceleration parameter in the series expansion of the distance--redshift relation, and which can be directly measured with upcoming cosmological datasets of standard candles. Thus, we argue that the formalism presented in the present paper is the relevant for the analysis of cosmological surveys of standardisable objects. } 

{Another insight of this analysis is that the effective observational parameters replacing the FLRW Hubble, deceleration, jerk, and curvature parameters in the generalised Hubble law are anisotropic in a particular way which can be directly linked to kinematic and dynamic properties of the observer congruence.}  
The existence of a dipole in the observed deceleration parameter from peculiar velocity flow around an FLRW solution has been hypothesized in \cite{Tsagas:2009nh,Tsagas:2011wq}. In this paper we confirm the general expectation of a dipolar structure in the effective deceleration parameter of the luminosity distance Hubble law, though the deceleration parameter and its dipole takes a different form than that proposed in \cite{Tsagas:2009nh,Tsagas:2011wq}. 
As detailed in section \ref{sec:multipole}, the effective deceleration parameter $\mathfrak{Q}$ differs -- even in its monopole limit -- from the natural generalisation \sayy{$q$}$:= -1 -  3 \frac{{\rm d}   \theta }{{\rm d} \tau} / \theta^2$ of the FLRW deceleration parameter.
We also show that the effective deceleration parameter in general caries multipole terms up to fourth order, whereas the effective jerk parameter in general carries multipole terms up to sixth order.  
The derived series expansion can be generalised to higher orders in redshift, yielding coefficients with multipole expansions truncated at yet higher order. 

Some multipole studies of distance--redshift data have already been carried out \cite{Wang:2014vqa,Bolejko:2015gmk,Bernal:2016kfw,Colin:2018ghy}, which may be further developed using the results of this paper. 
Analyses determining the local expansion rate, which have so far been carried out with isotropic assumptions for the distance ladder \cite{Riess:2016jrr,Riess:2019cxk}, may also be adapted to the presented model independent framework. 
The results of the present paper can be seen as extending the results on power series expansions of the angular diameter distance in relativistic space-times and multipole formulations of these \cite{KristianSachs,EllisMacCallum,Clarkson:2011uk,Clarkson:2011br}. 
Another luminosity distance multipole analysis framework has been proposed for investigation of linearly perturbed FLRW space-times \cite{Bonvin:2005ps,Bonvin:2006en}. 
While the approach in \cite{Bonvin:2005ps,Bonvin:2006en} relies on a direct multipole expansion of the luminosity distance, our approach is based on a convenient form of the anisotropic coefficients $d_L^{(1)}(\bm{e})$, $d_L^{(2)}(\bm{e})$, and $d_L^{(3)}(\bm{e})$ of the series expansion of the luminosity distance in redshift, allowing us to express these in terms of a \emph{finite} set of multipole coefficients. 
This makes our procedure convenient for model independent analysis, invoking only the minimal set of assumptions (i)--(v) discussed in section \ref{sec:assumptions} such that $z \mapsto d_L$ is a single-valued function with a well defined taylor expansion for each null ray. 
While the framework presented in this paper is much broader than the perturbed FLRW subcase, it may be used in this limit as well, yielding a complementary approach to that of \cite{Bonvin:2005ps,Bonvin:2006en} in the low to intermediate redshift regime. 

A possible direction for future research based on the framework presented in this paper is the orientation of CMB multipoles relative to the multipoles of the luminosity distance Hubble law described in this work. Since the majority of multipole contributions arise from the reparametrization $\lambda \mapsto z(\lambda)$ of the null rays (the remaining contributions come directly from the focusing equation (\ref{focuseq}) -- see (\ref{dAexpand2})) one could hypothesise the entering of correlation between multipoles of the CMB and multipoles entering the luminosity distance Hubble law. 
Another potential direction for future research drawing on the results in this paper is the direct testing of the FLRW geometrical assumptions. 
The multipole contributions entering the distance expansion appear in extended versions of the Ehlers-Geren-Sachs theorem\footnote{The Ehlers-Geren-Sachs theorem \cite{EGS} and generalisations hereof \cite{EGSStoeger,Rasanen:2009mg} seek to constrain the level of homogeneity and isotropy of the space-time metric from the level of isotropy of the CMB temperature field as inferred by classes of observers. So far an issue for applying the theorems to data has been that they make assumptions on the \emph{derivatives} of the energy function, which are not directly observable in CMB data.} \cite{EGSStoeger,Rasanen:2009mg} that make assumptions on the relative order of magnitude of the kinematical variables $\theta$, $a^\mu$, $\sigma_{\mu \nu}$, and their derivatives along with the off-diagonal components of the Ricci tensor $h^{\alpha}_\mu u^\nu R_{\alpha \nu}$ and $R_{\la \mu \nu \ra }$ to prove the \sayy{closeness} of the cosmic space-time geometry to an FLRW metric\footnote{For a debate on the accuracy of the FLRW approximation on nested scales and the significance of cosmological backreaction, see \cite{Green:2014aga,Buchert:2015iva}.}. 
These terms appear directly in the observable multipole coefficients $\{ \Eu_o, \widehat{\mathfrak{Q}}_o, \widehat{\mathfrak{J}}_o \}$ -- see (\ref{dLexpandcoef}) or the explicit expressions for the multipole coefficients (\ref{Eu}), (\ref{qpoles}), (\ref{tpoles}) and compare with Theorem 2 of \cite{Rasanen:2009mg}. 
Model independent determination of the size of the multipole terms entering the general Hubble law may thus be used as a direct test of the FLRW geometric approximation.\footnote{A potential obstacle to such tests is the regularity assumption (iii) imposed in section \ref{sec:assumptions}, that effectively requires the consideration of scales large enough for the expansion rate $\theta$ to dominate over acceleration $a^\mu$ and shear $\sigma_{\mu \nu}$ contributions (as projected onto any spatial direction). However, the variables $\theta$, $a^\mu$, and $\sigma_{\mu \nu}$ can still consistently be of similar order of magnitude within the proposed framework.} 
Yet another direction of research inspired by the framework presented in this paper, is the model independent formulation of other observables exploiting the decomposition of physical observables into truncated multipole series. See, for instance, the multipole decomposition of redshift drift signals in general space-times presented in \cite{Heinesen:2020pms}. 

In conclusion we have formulated an expression for the luminosity distance Hubble law for general space-times under a minimal set of assumptions. 
The simple form in terms of physically interpretable multipole coefficients makes the expression directly applicable to data analysis with existing and upcoming surveys. 

\newpage 

\acknowledgments
This work is part of a project that has received funding from the European Research Council (ERC) under the European Union's Horizon 2020 research and innovation programme (grant agreement ERC advanced grant 740021--ARTHUS, PI: Thomas Buchert). I thank Thomas Buchert for his careful reading and comments on the manuscript.  I also thank Chris Clarkson and the anonymous referee for useful comments.

\begin{appendices} 

\section{Traceless reduction of symmetric tensors} 
\label{tracereduction}
A tensor in three dimensions can be expressed in therms of traceless symmetric tensors and isotropic tensors \cite{Spencer:1970}. 
We shall be interested in the decomposition of symmetric and spatially projected tensors satisfying $T_{\mu_1 , \mu_2, .. , \mu_n } = T_{(\mu_1 , \mu_2, .. , \mu_n )} = h_{ \mu_1 }^{\, \alpha_1 } h_{ \mu_2 }^{\, \alpha_2 } .. h_{ \mu_n }^{\, \alpha_n }     T_{(\alpha_1 , \alpha_2, .. , \alpha_n)}$, where $n$ denotes the number of indices. 
Here we provide the reduction of $T_{\mu_1 , \mu_2, .. , \mu_n }$ into traceless symmetric tensors for $n\leq 6$: 
\bea
\label{tracerules} 
T_{\la \mu \nu \ra}  &=& T_{\mu \nu} - \frac{1}{3}h_{\mu \nu} T \, ,  \nonumber \\
T_{\la \mu \nu \rho \ra}  &=& T_{\mu \nu \rho} - \frac{1}{5} \left( T_{\mu} h_{\nu \rho}  + T_{\nu} h_{\rho \mu} + T_{\rho} h_{\mu \nu} \right) \, ,   \nonumber \\
T_{\la \mu \nu \rho \kappa \ra}  &=& T_{\mu \nu \rho \kappa} - \frac{1}{7} \left( T_{ \la \mu \nu \ra} h_{\rho \kappa} + T_{ \la \mu \rho \ra} h_{\nu \kappa}  + T_{ \la \mu \kappa \ra} h_{\nu \rho}  + T_{ \la \nu \rho \ra} h_{\mu \kappa}  \right.  \nonumber \\ 
  &+& \left.   T_{ \la \nu \kappa \ra} h_{ \mu \rho} + T_{ \la  \rho \kappa \ra} h_{  \mu \nu}    \right)  -  \frac{1}{15}T H_{\mu \nu \rho \kappa} \, ,  \nonumber \\
  T_{\la \mu \nu \rho \kappa \gamma \ra}  &=&   T_{ \mu \nu \rho \kappa \gamma }  - \frac{1}{9} \left( T_{\la \mu \nu \rho \ra} h_{\kappa \gamma} + T_{\la \mu \nu \kappa \ra} h_{\rho \gamma} + T_{\la \mu \nu \gamma \ra} h_{\rho \kappa}   \right. \nonumber \\ 
  &+& T_{\la \mu \rho \kappa \ra} h_{\nu \gamma} + \left. T_{\la \mu \rho \gamma \ra} h_{\nu \kappa} + T_{\la \mu \kappa \gamma \ra} h_{\nu \rho}   + T_{\la \nu \rho \kappa \ra} h_{\mu \gamma}   \right. \nonumber \\ 
  &+& \left. T_{\la \nu \rho \gamma \ra} h_{\mu \kappa}  + T_{\la \nu \kappa \gamma \ra} h_{\mu \rho} + T_{\la \rho \kappa \gamma \ra} h_{\mu \nu} \right)  \nonumber \\ 
  &-& \frac{1}{35} \left(T_\mu H_{\nu \rho \kappa \gamma} + T_\nu H_{\mu \rho \kappa \gamma}   + T_\rho H_{\mu \nu \kappa \gamma} + T_\kappa H_{\mu \nu \rho \gamma} + T_\gamma H_{\mu \nu \rho \kappa}     \right) \,  \nonumber \\  
    T_{\la \mu \nu \rho \kappa \gamma \sigma \ra}  &=&   T_{ \mu \nu \rho \kappa \gamma \sigma } - \frac{1}{11} \left( T_{\la \mu \nu \rho \kappa  \ra} h_{\gamma \sigma} + T_{\la \mu \nu \rho \gamma  \ra} h_{\kappa \sigma}  + T_{\la \mu \nu \rho \sigma  \ra} h_{\kappa \gamma}  \right. \nonumber \\  
 &+& \left.  T_{\la \mu \nu  \kappa  \gamma \ra} h_{\rho \sigma}    +  T_{\la \mu \nu  \kappa  \sigma \ra} h_{\rho \gamma}   + T_{\la \mu \nu \gamma \sigma  \ra} h_{\rho \kappa} + T_{\la \mu \rho \kappa \gamma  \ra} h_{\nu \sigma}   \right. \nonumber \\  
 &+& \left.  T_{\la \mu \rho \kappa \sigma  \ra} h_{\nu \gamma}  +  T_{\la \mu \rho \gamma \sigma  \ra} h_{\nu \kappa}  + T_{\la \mu \kappa \gamma \sigma  \ra} h_{\nu \rho} +  T_{\la \nu \rho \kappa \gamma  \ra} h_{\mu \sigma}    \right. \nonumber \\  
 &+& \left.  T_{\la \nu \rho \kappa \sigma \ra} h_{\mu \gamma}  + T_{\la \nu \rho \gamma \sigma \ra} h_{\mu \kappa}  +   T_{\la \nu \kappa \gamma \sigma \ra} h_{\mu \rho} +   T_{\la \rho \kappa \gamma \sigma \ra} h_{\mu \nu} \right) \nonumber \\  
 &-&\frac{1}{63}  \left( T_{\la \gamma \sigma \ra}  H_{ \mu \nu \rho \kappa  } + T_{\la \kappa \sigma \ra } H_{\mu \nu \rho \gamma  }   + T_{\la \kappa \gamma \ra} H_{ \mu \nu \rho \sigma  }   \right. \nonumber \\  
 &+& \left.  T_{\la \rho \sigma \ra}   H_{ \mu \nu  \kappa  \gamma }   +  T_{\la \rho \gamma \ra }  H_{ \mu \nu  \kappa  \sigma }   + T_{\la \rho \kappa \ra} H_{ \mu \nu \gamma \sigma  }  + T_{\la \nu \sigma \ra}  H_{ \mu \rho \kappa \gamma  }   \right. \nonumber \\  
 &+& \left. T_{\la \nu \gamma \ra}  H_{ \mu \rho \kappa \sigma  }  +  T_{\la \nu \kappa \ra }  H_{ \mu \rho \gamma \sigma }  + T_{\la \nu \rho \ra } H_{ \mu \kappa \gamma \sigma  }  + T_{\la \mu \sigma \ra }  H_{ \nu \rho \kappa \gamma  }    \right. \nonumber \\  
 &+& \left.  T_{\la \mu \gamma \ra} H_{ \nu \rho \kappa \sigma }   + T_{\la \mu \kappa \ra} H_{ \nu \rho \gamma \sigma }   +   T_{\la \mu \rho \ra}  H_{ \nu \kappa \gamma \sigma } +  T_{\la \mu \nu \ra}  H_{ \rho \kappa \gamma \sigma } \right) \nonumber \\   
 &-& \frac{1}{7} T  h_{ ( \mu \nu}h_{\rho \kappa} h_{\gamma \sigma )}  \, , 
\eea 
where $H_{\mu \nu \rho \kappa} \equiv  3 h_{ ( \mu \nu}h_{\rho \kappa)}$, and where we have used the short hand notation for tensor contractions:  $T_{ \mu \nu \rho \kappa} \equiv h^{\gamma \sigma} T_{ \mu \nu \rho \kappa \gamma \sigma}$, $T_{ \mu \nu \rho} \equiv h^{\kappa \gamma} T_{ \mu \nu \rho \kappa \gamma}$, $T_{ \mu \nu} \equiv h^{\rho \kappa} T_{ \mu \nu \rho \kappa}$, $T_{ \mu} \equiv  h^{\nu \rho } T_{ \mu \nu \rho}$, and $T \equiv h^{\mu \nu} T_{\mu \nu}$. 
The projections of the reductions (\ref{tracerules}) onto a spatial unit vector field $\bm{e}$, lying within the 3--dimensional projection space $e^{\mu} = h^{\mu}_{\, \nu}e^\nu$, yield 
\bea
\label{traceproj} 
\hspace*{-0.8cm} && e^\mu \! e^\nu T_{\la \mu \nu \ra}  = e^\mu \! e^\nu T_{\mu \nu} - \frac{1}{3} T \, ,  \nonumber \\
\hspace*{-1cm} && e^\mu \! e^\nu \! e^\rho T_{\la \mu \nu \rho \ra}  = e^\mu \! e^\nu \! e^\rho T_{\mu \nu \rho} - \frac{3}{5} e^\mu T_{\mu}  \, ,   \nonumber \\
\hspace*{-1cm} && e^\mu \! e^\nu \! e^\rho \! e^\kappa  T_{\la \mu \nu \rho \kappa \ra}  = e^\mu \! e^\nu \! e^\rho \! e^\kappa T_{\mu \nu \rho \kappa} - \frac{6}{7} e^\mu \! e^\nu  T_{ \la \mu \nu \ra}   -  \frac{1}{5}T  \, ,  \nonumber \\
\hspace*{-1cm} &&e^\mu \! e^\nu \! e^\rho \! e^\kappa \! e^\gamma  T_{\la \mu \nu \rho \kappa \gamma \ra}  = e^\mu \! e^\nu \! e^\rho \! e^\kappa \! e^\gamma   T_{ \mu \nu \rho \kappa \gamma }  - \frac{10}{9} e^\mu \! e^\nu \! e^\rho T_{\la \mu \nu \rho \ra} - \frac{3}{7} e^\mu T_\mu  \,  \nonumber \\  
\hspace*{-1cm}  && e^\mu \! e^\nu \! e^\rho \! e^\kappa \! e^\gamma \! e^\sigma    T_{\la \mu \nu \rho \kappa \gamma \sigma \ra}  =  e^\mu \! e^\nu \! e^\rho \! e^\kappa \! e^\gamma \! e^\sigma  T_{ \mu \nu \rho \kappa \gamma \sigma } - \frac{15}{11} e^\mu \! e^\nu \! e^\rho \! e^\kappa  T_{\la \mu \nu \rho \kappa  \ra}  - \frac{5}{7} e^\mu \! e^\nu   T_{\la \mu \nu \ra}  -  \frac{1}{7} T   \, .
\eea

\section{Multipole expansions of the effective curvature and jerk parameters} 
\label{app:multipole}
We consider the effective curvature parameter of the luminosity distance Hubble law 
\bea
\label{R}
\mathfrak{R} (\bm e ) &=&  -  \frac{  \overset{0}{\mathfrak{r}}   +  \bm{e} \cdot  \bm{{\overset{1}{\mathfrak{r}}}}   +    \bm{e} \bm{e} \cdot  \bm{{\overset{2}{\mathfrak{r}}}}     +    \bm{e} \bm{e} \bm{e} \cdot  \bm{{\overset{3}{\mathfrak{r}}}}    +    \bm{e} \bm{e} \bm{e} \bm{e} \cdot  \bm{{\overset{4}{\mathfrak{r}}}}        }{\Eu^2(\bm e )}  \, , 
\eea 
where 
\bea
\label{Rpoles} 
&& \overset{0}{\mathfrak{r}} \equiv \overset{0}{\mathfrak{q}} + \frac{1}{2} (u^{\mu} u^\nu + \frac{1}{3}h^{\mu \nu} )  R_{\mu \nu}    \, , \qquad \quad \overset{1}{\mathfrak{r}}_\mu \equiv  \overset{1}{\mathfrak{q}}_{\mu}   -  h^{\alpha}_\mu u^\nu R_{\alpha \nu}   \, , \nonumber \\
&& \overset{2}{\mathfrak{r}}_{\mu \nu}  \equiv    \overset{2}{\mathfrak{q}}_{\mu \nu} + \frac{1}{2}  R_{\la \mu \nu \ra }  \,    , \qquad \quad    \overset{3}{\mathfrak{r}}_{\mu \nu \rho}  \equiv   \overset{3}{\mathfrak{q}}_{\mu \nu \rho}  \, ,   \qquad \quad \overset{4}{\mathfrak{r}}_{\mu \nu \rho \kappa}  \equiv   \overset{4}{\mathfrak{q}}_{\mu \nu \rho \kappa}  \, , 
\eea 
which can be obtained directly from the definition (\ref{paramseff}) and (\ref{q}). 
In a general relativistic fluid approximation $(u^{\mu \nu} + h^{\mu \nu}/3)R_{\mu \nu}$ represents the sum of restmass density and isotropic pressure, $h^{\alpha}_\mu u^\nu R_{\alpha \nu}$ represents the energy flux, and $R_{\la \mu \nu \ra }$ represents the anisotropic stress with respect to the frame of $\bm{u}$. 
The effective curvature parameter $\mathfrak{R}$ is an anisotropic function of dynamical properties of the photon congruence and the Ricci curvature of the space-time. 
We can consider the monopole limit, where we neglect all anisotropic contributions in (\ref{Eu}) and (\ref{R}), such that $\mathfrak{R} \rightarrow - 9 \overset{0}{\mathfrak{r}} /\theta^2$. 
Due to the presence of vorticity, shear and acceleration degrees of freedom, $\mathfrak{R}$ does not reduce to its natural extension from FLRW geometry \sayy{$\Omega_k$}$:=-(3/2){}^{(3)}R/\theta^2$ in this limit. Thus even when neglecting anisotropic contributions to the luminosity distance--redshift relation, the effecive curvature parameter $\mathfrak{R}$ does not directly measure the 3-curvature ${}^{(3)}R$ of the spatial section defined by $\bm{u}$. 
We next consider the jerk parameter, which is given by the expansion 
\bea
\label{jerk}
\mathfrak{J}(\bm e ) &=& 1 +   \frac{  \overset{0}{\mathfrak{j}}   +  \bm{e} \cdot  \bm{{\overset{1}{\mathfrak{j}}}}   +    \bm{e} \bm{e} \cdot  \bm{{\overset{2}{\mathfrak{j}}}}     +    \bm{e} \bm{e} \bm{e} \cdot  \bm{{\overset{3}{\mathfrak{j}}}}    +    \bm{e} \bm{e} \bm{e} \bm{e} \cdot  \bm{{\overset{4}{\mathfrak{j}}}}      +   \bm{e} \bm{e} \bm{e} \bm{e} \bm{e} \cdot  \bm{{\overset{5}{\mathfrak{j}}}}     +   \bm{e} \bm{e} \bm{e} \bm{e} \bm{e} \bm{e} \cdot  \bm{{\overset{6}{\mathfrak{j}}}}  }{\Eu^3(\bm e )}    
\eea 
in terms of the multipole coefficients
\bea
\label{jpoles}
&& \overset{0}{\mathfrak{j}}  \equiv   \frac{ {\rm d} \overset{0}{\mathfrak{q}}  }{{\rm d} \tau} + \theta \overset{0}{\mathfrak{q}}   - \frac{1}{3} D^{ \mu} \overset{1}{\mathfrak{q}}_{\mu }   - \frac{1}{3} a^{  \mu} \overset{1}{\mathfrak{q}}_{\mu }  - \frac{2}{3} \sigma^{ \mu \nu}   \overset{2}{\mathfrak{q}}_{\mu \nu}     + h^{\mu \nu}   h^{\rho \kappa}   \sigma_{ ( \mu \nu } \overset{2}{\mathfrak{q}}_{\rho \kappa )}      \, , \nonumber \\
&&  \overset{1}{\mathfrak{j}}_\mu \equiv - D_\mu \overset{0}{\mathfrak{q}}  - 3 a_\mu \overset{0}{\mathfrak{q}} + h^{\nu}_{\, \mu} \frac{ {\rm d}  \overset{1}{\mathfrak{q}}_{\nu}  }{{\rm d} \tau}  +  \theta \overset{1}{\mathfrak{q}}_{\mu}  - S^\nu_{\; \mu} \overset{1}{\mathfrak{q}}_{\nu}  
  + 2 a^\nu  \overset{2}{\mathfrak{q}}_{\mu \nu}     +  \frac{18}{7}  h^{\nu \rho}   h^{\kappa \gamma}    \sigma_{( \mu \nu }  \overset{3}{\mathfrak{q}}_{ \rho \kappa \gamma)}   \nonumber \\ 
  && \qquad \quad + \frac{1}{5} h^{\nu \rho} \left( 4  \sigma_{ (\mu \nu } \overset{1}{\mathfrak{q}}_{\rho )}   - D_{(\mu} \overset{2}{\mathfrak{q}}_{\nu \rho)}  -  5 a_{(\mu} \overset{2}{\mathfrak{q}}_{\nu \rho)}      -3  S^\kappa_{\; ( \mu}  \overset{3}{\mathfrak{q}}_{\nu \rho ) \kappa } \right)       \, , \nonumber \\
 && \overset{2}{\mathfrak{j}}_{\mu \nu}  \equiv   3  \overset{0}{\mathfrak{q}} \sigma_{\mu \nu}   - D_{ \la \mu} \overset{1}{\mathfrak{q}}_{\nu \ra }   - 4 a_{ \la \mu} \overset{1}{\mathfrak{q}}_{\nu \ra }  - 2 S^\rho_{\; \la \mu}   \overset{2}{\mathfrak{q}}_{\nu \ra \rho}    + h^{\alpha}_{\, \mu} h^{\beta}_{\, \nu } \frac{ {\rm d}  \overset{2}{\mathfrak{q}}_{\alpha \beta} }{{\rm d} \tau}  +   \theta \overset{2}{\mathfrak{q}}_{\mu \nu}           + 3 a^\rho \overset{3}{\mathfrak{q}}_{\mu \nu \rho}   \nonumber \\ 
 && \qquad \quad +   \frac{6}{7} h_{\la \mu}^{\, \alpha} h_{\nu \ra}^{\, \beta}   h^{\rho \kappa}  \left(  5 \sigma_{ ( \alpha \beta } \overset{2}{\mathfrak{q}}_{\rho \kappa )}     - D_{(\alpha}  \overset{3}{\mathfrak{q}}_{\beta \rho \kappa)}  -  6  a_{( \alpha}  \overset{3}{\mathfrak{q}}_{ \beta \rho \kappa )}  - 4 S^\gamma_{\; ( \alpha}  \overset{4}{\mathfrak{q}}_{\beta \rho \kappa ) \gamma  }        \right)  \,  \nonumber \\ 
  && \qquad \quad    +   5  h_{\la \mu}^{\, \alpha} h_{\nu \ra}^{\, \beta}   h^{\rho \kappa}  h^{\gamma \sigma} \sigma_{( \alpha \beta }  \overset{4}{\mathfrak{q}}_{ \rho \kappa \gamma \sigma ) } \, ,   \nonumber \\ 
&&  \overset{3}{\mathfrak{j}}_{\mu \nu \rho}  \equiv    4  \sigma_{ \la \mu \nu } \overset{1}{\mathfrak{q}}_{\rho \ra }   - D_{\la \mu} \overset{2}{\mathfrak{q}}_{\nu \rho \ra }  -  5 a_{\la \mu} \overset{2}{\mathfrak{q}}_{\nu \rho \ra }      -3  S^\kappa_{\; \la \mu}  \overset{3}{\mathfrak{q}}_{\nu \rho \ra  \kappa }           +  h^{\alpha}_{\, \mu} h^{\beta}_{\, \nu}   h^{\gamma}_{\, \rho}    \frac{ {\rm d}  \overset{3}{\mathfrak{q}}_{\alpha \beta \gamma}   }{{\rm d} \tau}   + \theta  \overset{3}{\mathfrak{q}}_{\mu \nu \rho}   \nonumber \\  
&& \qquad \quad   + 4 a^\kappa  \overset{4}{\mathfrak{q}}_{\mu \nu \rho \kappa} + \frac{10}{9} h_{\la \mu}^{\, \alpha} h_\nu^{\, \beta} h_{\rho \ra}^{\, \epsilon}   h^{\kappa \gamma}  \left(  6   \sigma_{( \alpha \beta }  \overset{3}{\mathfrak{q}}_{ \epsilon \kappa \gamma)} - D_{( \gamma} \overset{4}{\mathfrak{q}}_{\alpha \beta \epsilon \kappa ) }  -  7 a_{( \alpha } \overset{4}{\mathfrak{q}}_{ \beta \epsilon \kappa \gamma )} \right)   \, , \nonumber \\ 
&&   \overset{4}{\mathfrak{j}}_{\mu \nu \rho \kappa}  \equiv  5 \sigma_{ \la \mu \nu } \overset{2}{\mathfrak{q}}_{\rho \kappa \ra }  - D_{\la \mu}  \overset{3}{\mathfrak{q}}_{\nu \rho \kappa \ra}  - 6 a_{\la \mu}  \overset{3}{\mathfrak{q}}_{ \nu \rho \kappa \ra }  - 4 S^\gamma_{\; \la \mu}  \overset{4}{\mathfrak{q}}_{\nu \rho \kappa \ra \gamma  }   \nonumber  \\ 
&&\qquad \quad +  h^{\alpha}_{\,  \mu} h^{\beta}_{\, \nu  } h^{\sigma}_{\, \rho} h^{\eta}_{\, \kappa}   \frac{ {\rm d}    \overset{4}{\mathfrak{q}}_{\alpha \beta \sigma \gamma}   }{{\rm d} \tau}   + \theta \overset{4}{\mathfrak{q}}_{\mu \nu \rho \kappa}  + \frac{105}{11} h_{\la \mu}^{\, \alpha} h_\nu^{\, \beta} h_\rho^{\, \epsilon} h_{\kappa \ra}^{\, \psi}   h^{\gamma \sigma} \sigma_{( \alpha \beta }  \overset{4}{\mathfrak{q}}_{ \epsilon \psi \gamma \sigma ) }      \, , \nonumber \\ 
  &&   \overset{5}{\mathfrak{j}}_{\mu \nu \rho \kappa \gamma}  \equiv   6   \sigma_{\la \mu \nu }  \overset{3}{\mathfrak{q}}_{ \rho \kappa \gamma \ra} - D_{\la \gamma} \overset{4}{\mathfrak{q}}_{\mu \nu \rho \kappa \ra }  -  7 a_{\la \mu } \overset{4}{\mathfrak{q}}_{ \nu \rho \kappa \gamma \ra}    \, , \nonumber \\ 
&&   \overset{6}{\mathfrak{j}}_{\mu \nu \rho \kappa \gamma \sigma}  \equiv     7 \sigma_{\la \mu \nu }  \overset{4}{\mathfrak{q}}_{ \rho \kappa \gamma \sigma \ra } \, , 
\eea 
where $S^\alpha_{\; \nu} \equiv \sigma^\alpha_{\; \nu} + \omega^\alpha_{\; \nu}$. The notation $T_{\la \mu_1 , \mu_2, .. , \mu_n \ra}$ denotes the symmetric and traceless part of the spatial tensor $T_{\mu_1 , \mu_2, .. , \mu_n} =  h_{ \mu_1 }^{\, \alpha_1 } h_{ \mu_2 }^{\, \alpha_2 } .. h_{ \mu_n }^{\, \alpha_n }     T_{\alpha_1 , \alpha_2, .. , \alpha_n}$ as defined in appendix \ref{tracereduction}. 
Neglecting all anisotropic contributions in (\ref{Eu}) and (\ref{jerk}), we have $\mathfrak{J} \rightarrow 1 + 9 \overset{0}{\mathfrak{j}}/\theta^2$. 
Due to the presence of acceleration, shear, vorticity and spatial gradients in the expansion rate of the congruence, this limit does not correspond to the parameter \sayy{$j$}$:=1 + 9 ( \frac{{\rm d}^2   \theta }{{\rm d} \tau^2} + \theta \frac{{\rm d}   \theta }{{\rm d} \tau} ) / \theta^3 \rvert_{o}$ describing the  \sayy{isotropized jerk} in the frame of $\bm{u}$.  
Thus, the monopole limit of the effective jerk parameter of the general Hubble law does not directly measure the physical \sayy{volume jerk} associated with the observer congruence. 
We can finally consider the following useful combination of $\mathfrak{J}$ and $\mathfrak{R}$: 
\bea
\label{jerkminusR}
\hspace*{-0.4cm}\mathfrak{J}(\bm e ) - 1 - \mathfrak{R}(\bm e )  &=& \frac{  \overset{0}{\mathfrak{t}}   +  \bm{e} \cdot  \bm{{\overset{1}{\mathfrak{t}}}}   +    \bm{e} \bm{e} \cdot  \bm{{\overset{2}{\mathfrak{t}}}}     +    \bm{e} \bm{e} \bm{e} \cdot  \bm{{\overset{3}{\mathfrak{t}}}}    +    \bm{e} \bm{e} \bm{e} \bm{e} \cdot  \bm{{\overset{4}{\mathfrak{t}}}}      +   \bm{e} \bm{e} \bm{e} \bm{e} \bm{e} \cdot  \bm{{\overset{5}{\mathfrak{t}}}}     +   \bm{e} \bm{e} \bm{e} \bm{e} \bm{e} \bm{e} \cdot  \bm{{\overset{6}{\mathfrak{t}}}}  }{\Eu^3(\bm e )}    \, , 
\eea 
with  
\bea
\label{tpoles}
\hspace*{0cm}&& \overset{0}{\mathfrak{t}}  \equiv    \overset{0}{\mathfrak{j}}  + \frac{1}{3} \theta  \overset{0}{\mathfrak{r}}  + \frac{1}{5}  h^{\mu \nu} h^{\rho \kappa}  \sigma_{ ( \mu \nu } \overset{2}{\mathfrak{r}}_{\rho \kappa )}  - \frac{1}{3} a^{\mu}  \overset{1}{\mathfrak{r}}_{\mu}     \, , \nonumber \\ 
  &&  \overset{1}{\mathfrak{t}}_\mu \equiv \overset{1}{\mathfrak{j}}_\mu + \frac{1}{3} \theta  \overset{1}{\mathfrak{r}}_\mu  - a_\mu  \overset{0}{\mathfrak{r}}  +  \frac{3}{7} h^{\nu \rho} h^{\kappa \gamma}   \sigma_{( \mu \nu }  \overset{3}{\mathfrak{q}}_{ \rho \kappa \gamma)}    - \frac{3}{5} h^{\nu \rho} a_{(\mu}  \overset{2}{\mathfrak{r}}_{\nu \rho )} + \frac{3}{5} h^{\nu \rho} \sigma_{( \mu \nu}  \overset{1}{\mathfrak{r}}_{\rho )}      \, ,   \nonumber \\ 
 &&  \overset{2}{\mathfrak{t}}_{\mu \nu}  \equiv    \overset{2}{\mathfrak{j}}_{\mu \nu}  + \frac{1}{3} \theta  \overset{2}{\mathfrak{r}}_{\mu \nu}  - a_{\la \mu}  \overset{1}{\mathfrak{r}}_{\nu \ra } + \sigma_{\mu \nu}  \overset{0}{\mathfrak{r}}     + \frac{5}{7}h_{\la \mu}^{\, \alpha} h_{\nu \ra}^{\, \beta}   h^{\rho \kappa}  h^{\gamma \sigma} \sigma_{( \alpha \beta }  \overset{4}{\mathfrak{q}}_{ \rho \kappa \gamma \sigma ) }      \nonumber \\   
 &&  \qquad \; + \,   \frac{6}{7} h_{\la \mu}^{\, \alpha} h_{\nu \ra}^{\, \beta}   h^{\rho \kappa}  \left(  -  a_{( \alpha}  \overset{3}{\mathfrak{q}}_{ \beta \rho \kappa )}  +  \sigma_{ ( \alpha \beta } \overset{2}{\mathfrak{r}}_{\rho \kappa )}          \right)   \, , \nonumber \\ 
\hspace*{-1cm} &&  \overset{3}{\mathfrak{t}}_{\mu \nu \rho}  \equiv    \overset{3}{\mathfrak{j}}_{\mu \nu \rho}    + \frac{1}{3} \theta  \overset{3}{\mathfrak{q}}_{\mu \nu \rho}  - a_{\la \mu}  \overset{2}{\mathfrak{r}}_{\nu \rho \ra } + \sigma_{\la \mu \nu}  \overset{1}{\mathfrak{r}}_{\rho \ra }   +  \frac{10}{9}  h_{\la \mu}^{\, \alpha} h_\nu^{\, \beta} h_{\rho \ra}^{\, \epsilon}   h^{\kappa \gamma}  \left(   -   a_{( \alpha } \overset{4}{\mathfrak{q}}_{ \beta \epsilon \kappa \gamma )}  + \sigma_{( \alpha \beta }  \overset{3}{\mathfrak{q}}_{ \epsilon \kappa \gamma)}   \right) \, , \nonumber \\ 
\hspace*{-1cm} &&   \overset{4}{\mathfrak{t}}_{\mu \nu \rho \kappa}  \equiv   \overset{4}{\mathfrak{j}}_{\mu \nu \rho \kappa}     + \frac{1}{3} \theta  \overset{4}{\mathfrak{q}}_{\mu \nu \rho \kappa}  - a_{\la \mu}  \overset{3}{\mathfrak{q}}_{\nu \rho \kappa \ra } + \sigma_{\la \mu \nu}  \overset{2}{\mathfrak{r}}_{\rho \kappa \ra }   + \frac{15}{11} h_{\la \mu}^{\, \alpha} h_\nu^{\, \beta} h_\rho^{\, \epsilon} h_{\kappa \ra}^{\, \psi}   h^{\gamma \sigma} \sigma_{( \alpha \beta }  \overset{4}{\mathfrak{q}}_{ \epsilon \psi \gamma \sigma ) }    \, , \nonumber \\ 
\hspace*{-1cm}  &&   \overset{5}{\mathfrak{t}}_{\mu \nu \rho \kappa \gamma}  \equiv   \overset{5}{\mathfrak{j}}_{\mu \nu \rho \kappa \gamma}     - a_{\la \mu}  \overset{4}{\mathfrak{q}}_{\nu \rho \kappa \gamma \ra} + \sigma_{\la \mu \nu}  \overset{3}{\mathfrak{q}}_{\rho \kappa \gamma \ra }             \, ,  \nonumber \\ 
&&\overset{6}{\mathfrak{t}}_{\mu \nu \rho \kappa \gamma \sigma}  \equiv    \frac{8}{7} \overset{6}{\mathfrak{j}}_{\mu \nu \rho \kappa \gamma \sigma}        \, , 
\eea 
which follows directly from combing (\ref{R}) and (\ref{jerk}).

\end{appendices}



\begin{thebibliography}{99}

\bibitem{Lemaitre:1927}
G.~{Lema\^itre},
\emph{Un Univers homog\`ene de masse constante et de rayon croissant rendant compte de la vitesse radiale des n\'ebuleuses extra-galactiques},  
Annales de la Soci\'et\'e Scientifique de Bruxelles \href{https://ui.adsabs.harvard.edu/abs/1927ASSB...47...49L/abstract}{\textbf{A47} (1927), 49}, 
Republished in: Mon. Not. Roy. Astron. Soc. \href{https://ui.adsabs.harvard.edu/abs/1931MNRAS..91..483L/abstract}{\textbf{91} (1931), 438}  

\bibitem{Slipher:1917}
V.~M.~{Slipher},
\emph{Nebulae},  
Proceedings of the American Philosophical Society \href{https://ui.adsabs.harvard.edu/abs/1917PAPhS..56..403S/abstract}{\textbf{56} (1917), 403} 

\bibitem{Hubble:1929}
E.~P.~{Hubble},
\emph{A relation between distance and radial velocity among extra-galactic nebulae},  
PNAS \href{https://doi.org/10.1073/pnas.15.3.168}{\textbf{15} (1929), 168} 

\bibitem{Visser:2003vq}
M.~Visser,
\emph{Jerk, snap, and the cosmological equation of state}, 
Class. Quant. Grav. \href{https://doi.org/10.1088/0264-9381/21/11/006}{\textbf{21} (2004), 2603}    
[\href{https://arxiv.org/abs/gr-qc/0309109}{arXiv:gr-qc/0309109}]  

\bibitem{Celerier:1999hp}
M.~N.~Celerier,
\emph{Do we really see a cosmological constant in the supernovae data?}, 
Astron. Astrophys. \href{https://ui.adsabs.harvard.edu/abs/2000A\%26A...353...63C/abstract}{\textbf{353} (2000), 63}   
[\href{https://arxiv.org/abs/astro-ph/9907206}{arXiv:astro-ph/9907206}]  

\bibitem{Tanimoto:2007dq}
M.~Tanimoto and Y.~Nambu,
\emph{Luminosity distance-redshift relation for the LTB solution near the center}, 
Class. Quant. Grav. \href{https://doi.org/10.1088/0264-9381/24/15/006}{\textbf{24} (2007), 3843}   
[\href{https://arxiv.org/abs/gr-qc/0703012}{arXiv:gr-qc/0703012}]  

\bibitem{Villani:2014zta}
M.~Villani,
\emph{Taylor expansion of luminosity distance in Szekeres cosmological models: Effects of local structures evolution on cosmographic parameters}, 
JCAP  \href{https://doi.org/10.1088/1475-7516/2014/06/015}{\textbf{06} (2014), 015}  
[\href{https://arxiv.org/abs/1401.4820}{arXiv:1401.4820}]  

\bibitem{Seitz:1994xf}
Seitz~S, Schneider~P and Ehlers~J, 
\emph{Light propagation in arbitrary space-times and the gravitational lens approximation},  
Class. Quant. Grav. \href{https://doi.org/10.1088/0264-9381/11/9/016}{\textbf{11} (1994), 2345}  
[\href{https://arxiv.org/abs/astro-ph/9403056}{arXiv:astro-ph/9403056}]  

\bibitem{KristianSachs}
J.~Kristian and R.~K.~Sachs, 
\emph{Observations in cosmology}, 
Astrophysical Journal \href{https://doi.org/10.1086/148522}{\textbf{143} (1966), 379}   

\bibitem{EllisMacCallum}
G.~F.~R.~Ellis and M.~A.~H.~MacCallum, 
\emph{A class of homogeneous cosmological models. II. Observations}, 
Comm. Math. Phys. \href{https://projecteuclid.org/euclid.cmp/1103842615}{\textbf{19} (1970), 31}   

\bibitem{Clarkson:2011uk}
C.~Clarkson and O.~Umeh,
\emph{Is backreaction really small within concordance cosmology?}, 
Class. Quant. Grav. \href{https://doi.org/10.1088/0264-9381/28/16/164010}{\textbf{28} (2011), 164010}   
[\href{https://arxiv.org/abs/1105.1886}{arXiv:1105.1886}]  

\bibitem{Clarkson:2011br}
C.~Clarkson, G.~F.~R.~Ellis, A.~Faltenbacher, R.~Maartens, O.~Umeh and J.~P.~Uzan,
\emph{(Mis-)Interpreting supernovae observations in a lumpy universe}, 
Mon. Not. Roy. Astron. Soc. \href{https://doi.org/10.1111/j.1365-2966.2012.21750.x}{\textbf{426} (2012), 1121}   
[\href{https://arxiv.org/abs/1109.2484}{arXiv:1109.2484}]  


\bibitem{DyerRoeder:1974}
C.~C.~Dyer and R.~C.~Roeder, 
\emph{Observations in locally inhomogeneous cosmological models}, 
Astrophys. J. \href{https://doi.org/10.1086/152784}{\textbf{189} (1974), 167}   

\bibitem{Sasaki:1987}
M.~Sasaki,
\emph{The magnitude-redshift relation in a perturbed Friedmann universe}, 
Mon. Not. Roy. Astron. Soc. \href{https://doi.org/10.1093/mnras/228.3.653}{\textbf{228} (1987), 653}  

\bibitem{FutamaseSasaki:1989}
T.~Futamase and M.~Sasaki, 
\emph{Light propagation and the distance redshift relation in a realistic inhomogeneous universe}, 
Phys. Rev. D \href{https://doi.org/10.1103/PhysRevD.40.2502}{\textbf{189} (1989), 2502}   

\bibitem{Kantowski:1998}
R.~Kantowski, 
\emph{The effects of inhomogeneities on evaluating the mass parameter $\Omega_m$ and the cosmological constant $\Lambda$}, 
Astrophys. J. \href{https://doi.org/10.1086/306355}{\textbf{507} (1998), 483}   
[\href{https://arxiv.org/abs/astro-ph/9802208}{arXiv:astro-ph/9802208}]  

\bibitem{Sugiura:1999}
N.~Sugiura, N.~Sugiyama, M.~Sasaki, 
\emph{Anisotropies in luminosity distance},  
Progress of Theoretical Physics \href{https://doi.org/10.1143/PTP.101.903 }{\textbf{101} (1999), 903}   

\bibitem{PyneBirkinshaw:2004}
T.~Pyne and M.~Birkinshaw, 
\emph{The luminosity distance in perturbed FLRW space--times},  
Mon. Not. R. Astron. Soc. \href{https://doi.org/10.1111/j.1365-2966.2004.07362.x}{\textbf{348} (2004), 581}   
[\href{https://arxiv.org/abs/astro-ph/0310841}{arXiv:astro-ph/0310841}]  

\bibitem{Bonvin:2005ps}
C.~Bonvin, R.~Durrer and M.~A.~Gasparini,
\emph{Fluctuations of the luminosity distance}, 
Phys. Rev. D \href{https://doi.org/10.1103/PhysRevD.85.029901}{\textbf{73} (2006), 023523}   
[\href{https://arxiv.org/abs/astro-ph/0511183}{arXiv:astro-ph/0511183}]  

\bibitem{Fleury:2014gha}
P.~Fleury,
\emph{Swiss-cheese models and the Dyer-Roeder approximation}, 
JCAP \href{https://doi.org/10.1088/1475-7516/2014/06/054}{\textbf{06} (2014), 054}  
[\href{https://arxiv.org/abs/1402.3123}{arXiv:1402.3123}]  

\bibitem{Bentivegna:2016fls}
E.~Bentivegna, M.~Korzy\'nski, I.~Hinder and D.~Gerlicher,
\emph{Light propagation through black-hole lattices}, 
JCAP \href{https://doi.org/10.1088/1475-7516/2017/03/014}{\textbf{03} (2017), 014}   
[\href{https://arxiv.org/abs/1611.09275}{arXiv:1611.09275}]  

\bibitem{SikoraGlod}
Sikora~S and G\l\'od~K, 
\emph{Example of an inhomogeneous cosmological model in the context of backreaction}, 
Phys. Rev. D \href{https://doi.org/10.1103/PhysRevD.95.063517}{\textbf{95} (2017), 063517} 
[\href{https://arxiv.org/abs/1612.03604}{arXiv:1612.03604}] 

\bibitem{Wang:2014vqa}
J.~S.~Wang and F.~Y.~Wang,
\emph{Probing the anisotropic expansion from supernovae and GRBs in a model-independent way}, 
Mon. Not. Roy. Astron. Soc. \href{https://doi.org/10.1093/mnras/stu1279}{\textbf{443} (2014), 1680} 
[\href{https://arxiv.org/abs/1406.6448}{arXiv:1406.6448}] 

\bibitem{Bolejko:2015gmk}
K.~Bolejko, M.~A.~Nazer and D.~L.~Wiltshire,
\emph{Differential cosmic expansion and the Hubble flow anisotropy}, 
JCAP \href{https://doi.org/10.1088/1475-7516/2016/06/035}{\textbf{06} (2016), 035}   
[\href{https://arxiv.org/abs/1512.07364}{arXiv:1512.07364}]  

\bibitem{Bernal:2016kfw}
C.~Bernal, V.~H.~Cardenas and V.~Motta,
\emph{Asymmetry in the reconstructed deceleration parameter}, 
Phys. Lett. B \href{https://doi.org/10.1016/j.physletb.2016.12.008}{\textbf{765} (2017), 163}    
[\href{https://arxiv.org/abs/1606.07333}{arXiv:1606.07333}]  

\bibitem{Colin:2018ghy}
J.~Colin, R.~Mohayaee, M.~Rameez and S.~Sarkar, 
\emph{Evidence for anisotropy of cosmic acceleration}, 
Astron. Astrophys. \href{https://doi.org/10.1051/0004-6361/201936373}{\textbf{631} (2019), L13}    
[\href{https://arxiv.org/abs/1808.04597}{arXiv:1808.04597}]  

\bibitem{Hogg}
Hogg~D et al.,
\emph{Cosmic homogeneity demonstrated with luminous red galaxies}, 
Astrophys. J. \href{https://doi.org/10.1086/429084}{\textbf{624} (2005), 54}   
[\href{https://arxiv.org/abs/astro-ph/0411197}{arXiv:astro-ph/0411197}] 

\bibitem{Scrimgeour}
Scrimgeour~M et al.,
\emph{The WiggleZ Dark Energy Survey: the transition to large-scale cosmic homogeneity}, 
Mon. Not. R. Astr. Soc. \href{https://doi.org/10.1111/j.1365-2966.2012.21402.x}{\textbf{425} (2012), 116}   
[\href{https://arxiv.org/abs/1205.6812}{arXiv:1205.6812}] 

\bibitem{ShapleyAmes}
H.~Shapley and A.~Ames,
\emph{A survey of the external galaxies brighter than the thirteenth magnitude}, 
Annals of Harvard College Observatory \href{https://ui.adsabs.harvard.edu/abs/1932AnHar..88...41S/abstract}{\textbf{88} (1932), 41}   

\bibitem{Proust:2005jt}
D.~Proust et al., 
\emph{Structure and dynamics of the shapley supercluster}, 
Astron. Astrophys. \href{https://doi.org/10.1051/0004-6361:20052838}{\textbf{447} (2006), 133}    
[\href{https://arxiv.org/abs/astro-ph/0509903}{arXiv:astro-ph/0509903}] 

\bibitem{Gott:2003pf}
J.~R.~Gott et al.,
\emph{A map of the universe}, 
Astrophys. J. \href{https://doi.org/10.1086/428890}{\textbf{624} (2005), 463}   
[\href{https://arxiv.org/abs/astro-ph/0310571}{arXiv:astro-ph/0310571}]  

\bibitem{Aihara}
H.~Aihara et al.,
\emph{The eighth data release of the Sloan digital sky survey: First Data from SDSS-III},  
The Astrophysical Journal Supplement \href{https://doi.org/10.1088/0067-0049/193/2/29}{\textbf{193} (2011), 17}   
[\href{https://arxiv.org/abs/1101.1559}{arXiv:1101.1559}]  

\bibitem{Feindt}
Feindt~U et al.,
\emph{Measuring cosmic bulk flows with type Ia Supernovae from the Nearby Supernova Factory}, 
Astron. Astrophys. \href{https://doi.org/10.1051/0004-6361/201321880}{\textbf{560} (2013), A90}   
[\href{https://arxiv.org/abs/1310.4184}{arXiv:1310.4184}] 

\bibitem{Magoulas}
Magoulas~C et al.,
\emph{Measuring the cosmic bulk flow with 6dFGSv},  
Proc. IAU Symp. \href{https://doi.org/10.1017/S1743921316010115}{\textbf{308} (2016), 336}   


\bibitem{sac1961}
R.~Sachs,
\emph{Gravitational waves in general relativity, VI. The outgoing radiation condition},
Proc. R. Soc. Lond. A \href{http://dx.doi.org/10.1098/rspa.1961.0202}{\textbf{264},
(1961), 309} 

\bibitem{wal1984}
R.~M.~Wald,
\emph{General Relativity}, (University
of Chicago Press, Chicago, 1984) 

\bibitem{ell1971}
G.~F.~R.~Ellis,
\emph{Relativistic cosmology}, Proc. Int. School of
Physics ``Enrico Fermi" (Varenna), Course XLVII, edited by
R.~K.~Sachs (Academic Press, New York, 1971), 104--182.
Reprinted:
Gen. Relativ. Grav. \href{http://dx.doi.org/10.1007/s10714-009-0760-7}{\textbf{41} (2009), 581}. 

\bibitem{Perlick:2010zh}
V.~Perlick,
\emph{Gravitational lensing from a spacetime perspective}, 
[\href{http://arxiv.org/abs/1010.3416}{arXiv:1010.3416}] 

\bibitem{eth1933}
I.~M.~H.~Etherington,
\emph{On the definition of distance in general relativity},
Phil. Mag. J. Sci. {\bf 15}, 761--773 (1933).
Reprinted:
Gen. Relativ. Grav. \href{http://dx.doi.org/10.1007/s10714-007-0447-x}{\textbf{39} 
 (2007), 1055}

\bibitem{ellhve1999}
G.~F.~R.~Ellis and H.~van Elst,
\emph{Cosmological models} (Carg\`ese Lectures 1998),
in {Proc. of the NATO Advanced Study Institute on Theoretical
and Observational Cosmology, Carg\`ese, France, August 17--29, 1998\/}, 
edited by  M.~Lachi\`eze--Rey, (Kluwer Academic, Boston, 1999), 1--116
and NATO Science Series C {\bf 541}, 1--116 (1999).
[\href{http://arxiv.org/abs/gr-qc/9812046}{arXiv:gr-qc/9812046v5}].

\bibitem{Fleury:2015rwa}
P.~Fleury, J.~Larena and J.~P.~Uzan,
\emph{The theory of stochastic cosmological lensing}, 
JCAP \href{https://doi.org/10.1088/1475-7516/2015/11/022}{\textbf{11} (2015), 022}  
[\href{https://arxiv.org/abs/1508.07903}{arXiv:1508.07903}]  

\bibitem{Tsagas:2009nh}
C.~G.~Tsagas,
\emph{Large-scale peculiar motions and cosmic acceleration}, 
Mon. Not. Roy. Astron. Soc. \href{https://doi.org/10.1111/j.1365-2966.2010.16460.x}{\textbf{405} (2010), 503}   
[\href{https://arxiv.org/abs/0902.3232}{arXiv:0902.3232}]  

\bibitem{Tsagas:2011wq}
C.~G.~Tsagas,
\emph{Peculiar motions, accelerated expansion and the cosmological axis}, 
Phys. Rev. D \href{https://doi.org/10.1103/PhysRevD.84.063503}{\textbf{84} (2011), 063503}    
[\href{https://arxiv.org/abs/1107.4045}{arXiv:1107.4045}]  

\bibitem{Tsagas:2015mua}
C.~G.~Tsagas and M.~I.~Kadiltzoglou,
\emph{Deceleration parameter in tilted Friedmann universes}, 
Phys. Rev. D \href{https://doi.org/10.1103/PhysRevD.92.043515}{\textbf{92} (2015), 043515}   
[\href{https://arxiv.org/abs/1507.04266}{arXiv:1507.04266}]  

\bibitem{Reisenegger:2000vc}
A.~Reisenegger, H.~Quintana, E.~R.~Carrasco and J.~Maze,
\emph{The shapley supercluster. 3. Collapse dynamics and mass of the central concentration}, 
Astron. J. \href{https://doi.org/10.1086/301477}{\textbf{120} (2000), 523} 
[\href{https://arxiv.org/abs/astro-ph/0007211}{arXiv:astro-ph/0007211}]  

\bibitem{Pearson:2014hja}
D.~W.~Pearson, M.~Batiste and D.~J.~Batuski,
\emph{The largest gravitationally bound structures: the Corona Borealis supercluster -- mass and bound extent}, 
Mon. Not. Roy. Astron. Soc. \href{https://doi.org/10.1093/mnras/stu693}{\textbf{441} (2014), 1601}   
[\href{https://arxiv.org/abs/1404.1308}{arXiv:1404.1308}]  

\bibitem{Einasto:2016rhe}
M.~Einasto et al., 
\emph{Sloan Great Wall as a complex of superclusters with collapsing cores}, 
Astron. Astrophys. \href{https://doi.org/10.1051/0004-6361/201628567}{\textbf{595} (2016), A70}   
[\href{https://arxiv.org/abs/1608.04988}{arXiv:1608.04988}]  

\bibitem{Secrest:2020has} 
N.~Secrest, S.~von Hausegger, M.~Rameez, R.~Mohayaee, S.~Sarkar and J.~Colin,
\emph{A test of the cosmological principle with quasars}, 
Astrophys. J. Lett. \href{https://doi.org/10.3847/2041-8213/abdd40}{\textbf{908} (2021), L51}   
[\href{https://arxiv.org/abs/2009.14826}{arXiv:2009.14826}]  

\bibitem{Korzynski:2019oal}
M.~Korzy\'nski and E.~Villa,
\emph{Geometric optics in relativistic cosmology: new formulation and a new observable}, 
Phys. Rev. D \href{https://doi.org/10.1103/PhysRevD.101.063506}{\textbf{101} (2020), 063506}   
[\href{https://arxiv.org/abs/1912.04988}{arXiv:1912.04988}]  

\bibitem{Scolnic:2019apa}
D.~Scolnic et al.,
\emph{The next generation of cosmological measurements with type Ia supernovae}, 
[\href{https://arxiv.org/abs/1903.05128}{arXiv:1903.05128}]  

{
\bibitem{buchert:grgdust}
T.~Buchert,
\emph{On average properties of inhomogeneous fluids in general
relativity: dust cosmologies}, 
Gen. Relativ. Gravit. \href{http://dx.doi.org/10.1023/A:1001800617177}{{\bf 32},
105 (2000)}.
[\href{http://arxiv.org/abs/gr-qc/9906015}{arXiv:gr-qc/9906015}]

\bibitem{buchert:grgfluid}
T.~Buchert,
\emph{On average properties of inhomogeneous fluids in general relativity:
perfect fluid cosmologies}, 
Gen. Relativ. Gravit. \href{http://dx.doi.org/10.1023/A:1012061725841}{{\bf 33},
1381 (2001)}.
[\href{http://arxiv.org/abs/gr-qc/0102049}{arXiv:gr-qc/0102049}]
} 

\bibitem{Rasanen:2003fy}
S.~R\"as\"anen,
\emph{Dark energy from backreaction}, 
JCAP \href{http://dx.doi.org/10.1088/1475-7516/2004/02/003}{\textbf{02} (2004), 003}.
[\href{http://arxiv.org/abs/astro-ph/0311257}{arXiv:astro-ph/0311257}] 

\bibitem{Rasanen:2006kp}
S.~R\"as\"anen,
\emph{Accelerated expansion from structure formation}, 
JCAP \href{http://dx.doi.org/10.1088/1475-7516/2006/11/003}{\textbf{11} (2006), 003}.
[\href{http://arxiv.org/abs/astro-ph/0607626}{arXiv:astro-ph/0607626}] 

\bibitem{Riess:2016jrr}
A.~G.~Riess et al., 
\emph{A 2.4\% determination of the local value of the Hubble constant}, 
Astrophys. J. \href{https://doi.org/10.3847/0004-637X/826/1/56}{\textbf{826} (2016), 56}   
[\href{https://arxiv.org/abs/1604.01424}{arXiv:1604.01424}]  

\bibitem{Riess:2019cxk}
A.~G.~Riess, S.~Casertano, W.~Yuan, L.~M.~Macri and D.~Scolnic,
\emph{Large Magellanic Cloud cepheid standards provide a 1\% foundation for the determination of the Hubble constant and stronger evidence for physics beyond $\Lambda$CDM}, 
Astrophys. J. \href{https://doi.org/10.3847/1538-4357/ab1422}{\textbf{876} (2019), 85}   
[\href{https://arxiv.org/abs/1903.07603}{arXiv:1903.07603}]  

\bibitem{Bonvin:2006en}
C.~Bonvin, R.~Durrer and M.~Kunz,
\emph{The dipole of the luminosity distance: a direct measure of H(z)}, 
Phys. Rev. Lett. \href{https://doi.org/10.1103/PhysRevLett.96.191302}{\textbf{96} (2006), 191302}   
[\href{https://arxiv.org/abs/astro-ph/0603240}{arXiv:astro-ph/0603240}]  

\bibitem{EGSStoeger}
W.~R.~{Stoeger}, R.~{Maartens} and G.~F.~R.~{Ellis},  
\emph{Proving almost homogeneity of the universe: an almost Ehlers-Geren-Sachs theorem},  
Astrophys. J. \href{https://doi.org/10.1086/175496}{\textbf{443} (1995), 1}   

\bibitem{Rasanen:2009mg}
S.~R\"as\"anen,
\emph{On the relation between the isotropy of the CMB and the geometry of the universe}, 
Phys. Rev. D \href{https://doi.org/10.1103/PhysRevD.79.123522}{\textbf{79} (2009), 123522}   
[\href{https://arxiv.org/abs/0903.3013}{arXiv:0903.3013}]  

\bibitem{EGS}
J.~{Ehlers}, P.~{Geren} and R.~K.~{Sachs},  
\emph{Isotropic solutions of the Einstein-Liouville equations}, 
J. Math. Phys. \href{https://doi.org/10.1063/1.1664720}{\textbf{9} (1968), 1344}   

\bibitem{Green:2014aga}
S.~R.~Green and R.~M.~Wald,
\emph{How well is our universe described by an FLRW model?}, 
Class. Quant. Grav. \href{https://doi.org/10.1088/0264-9381/31/23/234003}{\textbf{31} (2014), 234003}   
[\href{https://arxiv.org/abs/1407.8084}{arXiv:1407.8084}]  

\bibitem{Buchert:2015iva}
T.~Buchert et al., 
\emph{Is there proof that backreaction of inhomogeneities is irrelevant in cosmology?}, 
Class. Quant. Grav. \href{https://doi.org/10.1088/0264-9381/32/21/215021}{\textbf{32} (2015), 215021}   
[\href{https://arxiv.org/abs/1505.07800}{arXiv:1505.07800}]  

\bibitem{Spencer:1970}
A.~J.~M.~Spencer, 
\emph{A note on the decomposition of tensors into traceless symmetric tensors}, 
Phys. Rev. D \href{https://doi.org/10.1103/PhysRevD.79.123522}{\textbf{79} (2009), 123522}   
[\href{https://arxiv.org/abs/0903.3013}{arXiv:0903.3013}]  

\bibitem{Heinesen:2020pms}
A.~Heinesen, 
\emph{Multipole decomposition of redshift drift: Model-independent mapping of the expansion history of the Universe}, 
Phys. Rev. D \href{https://doi.org/10.1103/PhysRevD.103.023537}{\textbf{103} (2021), 023537} 
[\href{http://arxiv.org/abs/2011.10048}{arXiv:2011.10048}]  





\end{thebibliography}
\end{document}